# Characterizing the mode-choice behavior in developing countries

Kathleen Salazar[a,b*], Jesús D. Díaz[a], Isabel C. García[c]

[a] *Department of Civil and Industrial Engineering, Pontificia Universidad Javeriana, Cali, Colombia;* [b] *Department of Computer and Decision Sciences. Universidad Nacional de Colombia, Medellín, Colombia;* [c] *Department of Natural Sciences and Mathematics, Pontificia Universidad Javeriana, Cali, Colombia;* [d] *Department of Economics, Pontificia Universidad Javeriana, Cali, Colombia*

\* kathleen.salazar@javerianacali.edu.co

Provide short biographical notes on all contributors here if the journal requires them.

# Characterizing the transport mode-choice behavior in developing countries

Urban mobility in developing countries, particularly in cities like Cali, Colombia, faces multifaceted challenges influenced by socioeconomic factors and the distinct characteristics of transport users Despite motorcycles emerging as a prevalent mode, their impact remains underexplored in the literature. This study employs extensive survey data collection, descriptive and cluster analysis, enabling an understanding of mode choice behaviors. Comparative analyses reveal significant relationships between transport choices and variables such as socio-economic status and age. Cluster analysis unveiled distinct behavior patterns among Cali's urban transport users. Identifying groups with similar preferences and perceptions enhanced comprehension of diverse user needs, helping to understand travel behavior within the urban transport context. This study contributes to close knowledge gap insight into the intricate landscape of urban transport in developing countries. The findings offer a foundation for informed policy development, emphasizing the importance of considering socioeconomic factors, travel preferences, and the prevalence of motorcycles in crafting effective urban strategies.

**Introduction**

Passenger transportation holds broad significance, attracting attention from both the public and private spheres. In countries situated in the southern hemisphere, public transportation systems frequently grapple with limitations in both quality and coverage (Hagen et al., 2016; Vasconcellos, 2013) Consequently, there has been a notable surge in the utilization of private vehicles, particularly motorcycles (Chiu & Guerra, 2023; Rodríguez et al., 2015). Two-wheelers have emerged as the preferred mode of transport for individuals with lower incomes engaged in labor activities (H.-L. Chang & Wu, 2008). However, the rapid proliferation of motorcycles on roadways has given rise to significant challenges, encompassing congestion, road accidents, and environmental pollution (Bakker, 2019; Suatmadi et al., 2019).

Despite the challenges they pose in terms of accidents and traffic chaos, motorcycles have garnered scant attention relative to other modes of transportation. Often perceived as undesirable, they are frequently overlooked, even when serving as the predominant mode of transport in the Global South (Salazar-Serna, Cadavid, et al., 2023; Wigan, 2002). Formulating strategies based on dynamics analyzed in developed countries may prove ineffective, as socioeconomic and cultural conditions profoundly influence human behavior and, consequently, the decision-making process in choosing transportation modes (Salazar-Serna, Hui, et al., 2023)

Given the need of devising policies tailored to specific geographical conditions and people's behavior, the comprehensive characterization of users across diverse transportation modes becomes of paramount importance. This understanding is essential for discerning the nuanced needs and preferences of the population, facilitating efficient planning, and enhancing the management of transportation systems (Chou, 1992). Recognizing the paucity of studies examining transport-mode choices including motorcycles, particularly within the context of the Global South, this study aims to address existing research gaps. This is the first work identifying the distinctive characteristics of urban transport users in developing countries with a specific emphasis on private motorcycles as a utilitarian travel alternative. The methodology employed involves data collection within a case study city in Colombia, followed by a process of descriptive and inferential data analysis, complemented by cluster analysis, to comprehensively characterize the mode choice behaviors of transport users.

Colombia, classified as a developing country by the World Bank, boasts a substantial motorcycle fleet exceeding 11 million units, comprising 63% of the total vehicular count (Asociación Nacional de Movilidad Sostenible ANDEMOS, 2023). This noteworthy prevalence not only exacerbates issues of vehicular congestion (Routledge, 2018), but also exerts a profound impact on environmental concerns, particularly in terms of pollution (Cuellar et al., 2016); motorcycles constitute the primary contributor to air pollution in the country, accounting for 53% of total transport emissions (Ministerio de transporte, 2022). Moreover, motorcycles exhibit the highest fatal accident rates with 60% of all recorded accidents involving them. Statistics reveal a total of 4,914 motorcycle-related deaths in 2022, resulting in a fatal rate per 100,000 population that stands among the highest in the region, as reported by the National Road Safety Agency (ANSV, 2023).

In Cali, one of Colombia's most densely populated cities and characterized by a substantial number of circulating motorcycles, the aforementioned challenges are notably pronounced. This urban center has witnessed the highest decline in public transit utilization across the country, experiencing an 85% decrease between 2007 and 2022 (DANE, 2023). Additionally, it holds the second-highest rates of both fatal and non-fatal accidents in Colombia, predominantly attributed to motorcyclists, according to data from the National Road Safety Agency (ANSV, 2023). The situation in Cali mirrors that of numerous other cities in developing countries sharing similar socioeconomic and built environment conditions. Findings from this study offer insights for policy makers, providing a basis for analyzing strategies to encourage the utilization of more extensive public transport systems and mitigating the externalities arising from the widespread use of motorcycles, all tailored to the specific characteristics of transport users.

The remainder of the paper is organized as follows: section 2 presents some background and related works; section 3 is divided into two subsections, the questionnaire and sample design for data collection and data analysis, which includes the descriptive and inferential analysis as well as the cluster analysis. Results and discussion about the differences among transport mode users are shown in section 4. Finally, section 5 draws conclusions, policy implications, and future work.

**Background and related work**

Cluster analysis serves as a statistical technique enabling the grouping of individuals into homogeneous clusters based on shared characteristics. The typology employed relies on a concept of similarity, where two individuals are considered closer if they share a greater number of common characteristics (Escofier & Pages, 1992). As highlighted by Clayton et al. (2020) the application of cluster analysis is instrumental in identifying and delineating distinct groups of individuals within a sample who exhibit similar perceptions and/or demographic attributes. This analytical approach proves valuable for comprehending the similarities and distinctions among various groups within the broader population. In the context of this study, it becomes particularly pertinent for urban transport users, whose mode choice behaviors exhibit significant variation. The exploration of these clusters is crucial for uncovering the underlying characteristics influencing the selection of specific transportation modes (Sifat et al., 2021).

Cluster analysis has been applied in diverse studies addressing mobility-related concerns. An illustrative example is the work of G. Chang et al. (1992) wherein survey data from 1005 respondents were employed to understand suburban commuting behavior by identifying key travelers' characteristics through the application of cluster analysis. The authors underscore the notable gap in existing studies pertaining to distinctions in travel behavior between trips from suburbs to the city center and suburb-to-suburb journeys. Findings indicate that the socio-economic context of each grouping appears to reasonably explain the pattern of displacement manifested. Employing multivariate cluster analysis proves instrumental in classifying and comprehending these patterns, offering a valuable tool for a more effective and nuanced approach to understanding suburban mobility.

Chou (1992) also utilized data from a previous transport survey, garnering 164 valid responses, to investigate the applicability of a common decision process to the mode-choice behavior of the majority of travelers. Employing cluster analysis on transportation attitudinal survey data, three homogeneous groups were identified, each characterized by distinctive decision-making patterns. This discernment implies the presence of diverse decision-making processes within travel mode-choice behavior.

In exploring the acceptance of fully automated vehicles (AVs) and individuals' willingness to share them, Clayton et al. (2020) conducted a survey with 899 participants. Employing cluster analysis, they scrutinized respondents' preferences, delved into associated demographic and psychosocial characteristics, and sought to unveil users' attitudes and preferences regarding the adoption and sharing of AVs. This investigation aimed to shed light on potential transformations in mobility practices.

Cluster formation operates on the premise that individuals with similar tendencies display comparable actions, while those with different tendencies exhibit contrasting behaviors. However, the formation of these clusters is conditioned by the chosen segmentation basis, encompassing the variables employed to identify homogeneous groups of individuals. In the realm of marketing research, common segmentation bases include socio-economic characteristics, personality traits, demographic variables, and specific product attributes, with some studies incorporating attitudinal elements as well (Chou, 1992).

Although there is a large number of clustering techniques and algorithms, the general process can be divided into the following stages: (1) data collection and cleaning, (2) representation, which implies the selection of a similarity measure and preparation of data to make it suitable for the algorithm, (3) clustering strategy that involves the choice of clustering algorithm, and (4) interpretation, this stage combines clustering results with other analyses to conclude. Clustering algorithms can be hierarchical, partitional, or spectral algorithms (Jin & Han, 2011). Partitional algorithms such as k-means or k-medoids are good at finding spherical or convex shaped clusters that do not contain excessive outliers or noise but fail when trying to identify arbitrary shapes because can easily get stuck to local optima. Hierarchical clustering is a wide-spread technique for its easy implementation and good visualization (through dendrograms). These algorithms are either agglomerative or divisive. Agglomerative algorithms start with each data point in its own cluster and iteratively merges the nearest pairs of clusters until a single cluster containing all data points is formed. Divisive algorithms start with all observations contained in the same cluster and iteratively splits until each observation forms an individual cluster (Rodrigo, 2023).

Agglomerative clustering algorithms are widespread-used for several reasons. First, the flexibility to identify natural patterns in the data without pre-specifying the number of groups is a notable strength. Additionally, the algorithm strives to minimize variance within groups while maximizing dissimilarity between them, ensuring consistency in grouping and revealing significant differences among resulting clusters. The method's ability to visually represent the data structure through a dendrogram is advantageous, offering a clear graphical hierarchy of clustering, depicting the order and distance of mergers between groups. Lastly, the algorithm's resilience against outliers and noise enhances its reliability, ensuring stable results even when confronted with irregularities or disturbances in survey data collection (Cebecauer et al., 2023).

Considering the advantages of the agglomerative clustering method, this paper uses data from a survey of transport users in Cali city to identify emerging groups and analyze variations among users of each transport mode though agglomerative clustering algorithms. This approach aims to uncover and comprehend the diverse needs and preferences of transport users in the case study city.

*Case Study: Cali*

Cali is a Colombian city located in the western part of the country and serves as the capital of Valle del Cauca state. It is the second most populous city after Bogotá and Medellín, with an estimated population of 2,319,684 inhabitants (DANE, 2020). The city covers an area of 619 km² and is divided into 22 districts. Furthermore, Cali is recognized as the hub of the metropolitan region, making it a pivotal point for economic, cultural, and social activities in the area.

Cali's public transportation system, known as the *Masivo Integrado de Occidente* (MIO), has been operational since 2008, featuring articulated buses, standard buses, feeder bus services serving various city zones, and a cable car. The articulated buses operate on dedicated lanes, encompassing over 50 stations along their routes. The MIO cable system spans four stations, covering approximately 2080 meters (MetroCali MIO, 2023). Public transportation includes urban collective options (buses and minibuses), rural vehicles adapted for passenger transport (campers), taxis, and public bicycles (CaliBici) (Alcaldía de Santiago de Cali, 2018).

Furthermore, Cali's road network incorporates segregated lanes for cyclists and pedestrians along certain city streets. Despite these provisions, the current reality reveals that the majority of transport users (70%) opt for private vehicles, diverging from public transport and non-motorized modes(Alcaldía de Santiago de Cali, 2022).

**Methods**

*Data collection*

This section provides an overview of the initial phase of the proposed methodology to understand the mobility dynamics of transportation users in Cali city.

*Sample design*

In the sample design, it is crucial to determine the required sample sizes to draw conclusions that are both reliable and representative of the area of interest. In this regard, a simple random sampling method was chosen to select participants, considering the working-age population of Cali. Participant selection was carried out through simple random sampling. The sample size calculation is based on a dichotomous variable, namely, gender. Following Krejcie and Morgan (1970) considering it as a dichotomous variable, let's assume an acceptable margin of error of 5%. Additionally, suppose the probability of committing a Type I error is also 5%. Cochran's formula (Cochran, 1977) with finite population correction states that the sample size is given by

$$n = \frac{n_0}{1+(n_0-1)/N} \qquad (1)$$

where

$$n_0 = \frac{t^2(p)(1-p)}{d^2} \qquad (2)$$

In general, *t* represents the t-value for α=0.025 in each tail of the distribution —i.e., *t*=1.96. Additionally, *p(1−p)* corresponds to the estimated variance, and d to the acceptable margin of error —i.e., *d*=0.05. By hypothesis, it is determined that *p*=1/2 with the purpose of obtaining the maximum estimated variance and, consequently, the maximum sample size. (Note that since the selected sample represents less than 5% of the population, the finite population correction factor of Cochran (1977) is not necessary).

*Survey design*

A survey was designed to gather information about factors that influence the mode choice of individuals residing in Cali city and its surrounding areas. Based on suggestions from the literature review we designed a questionnaire that asks anonymously questions about the travel behavior of urban commuters. Most of the questions are about revealed preference aspects. The survey was divided into sections as outlined in Table 1.

The first section includes questions about sociodemographic and socioeconomic data. Subsequently, respondents are asked about their primary mode of transportation, including details about travel habits such as average time and distance. Next, the users rate the factors influencing their choice; the factors include those considered in previous works (Agarwal et al., 2020; Ahmed et al., 2020; Faboya et al., 2020; Fan & Chen, 2020; Fu, 2021; Gadepalli et al., 2020; Ryan, 2020; Zapolskytė et al., 2020) as well as the results derived from focus groups conducted with transport users in the city.

The next section explores the social influence in the modal choice, asking about family and friends' approval and how much that input influences their choice. Finally, to understand people's perception of means of transportation the survey includes questions about motorcycles, private cars and public transit (BRT) which are the predominant modes in the city.

Table 1. Survey Sections.

| Section | Content |
|---------|---------|

| Sociodemographic information | Gender, age range, income level, occupation, location of residence, vehicles in the household. |
| --- | --- |
| Primary mode of transportation | Main mode of transportation, time period as user, satisfaction with the chosen mode, and reasons for choosing that mode. |
| Travel behavior | Average time and distance, frequent origin and destination, carpooling habits. |
| Influencing factors in choosing the transport mode | Level of importance users give to the transport attributes: Acquisition cost, maintenance expenses, road safety, personal security, comfort, travel time, and pollution. |
| Social influence | Level of approval from family and friends regarding the primary mode of transportation and how important their opinion is. |
| Perception of transport attributes | Opinions on costs, road safety, personal security, comfort, travel time, and pollution. Risk perception of being assaulted when using motorcycle, private cars, or public transit. |

*Survey Implementation*

The survey administration commenced in the month of February at various locations within Cali city using Google Forms, which gathers the responses into a centralized database. The questions were designed with the expectation that the primary respondents would be working-age individuals. The survey collection process concluded at the end of August. Throughout this period, the survey was administered continuously to collect a full and accurate range of responses. During the collection, emphasis was placed on institutions that would allow the survey to be administered around the southern part of the city, which is the focus of congestion. Figure 1 presents the collection points where the survey was implemented.

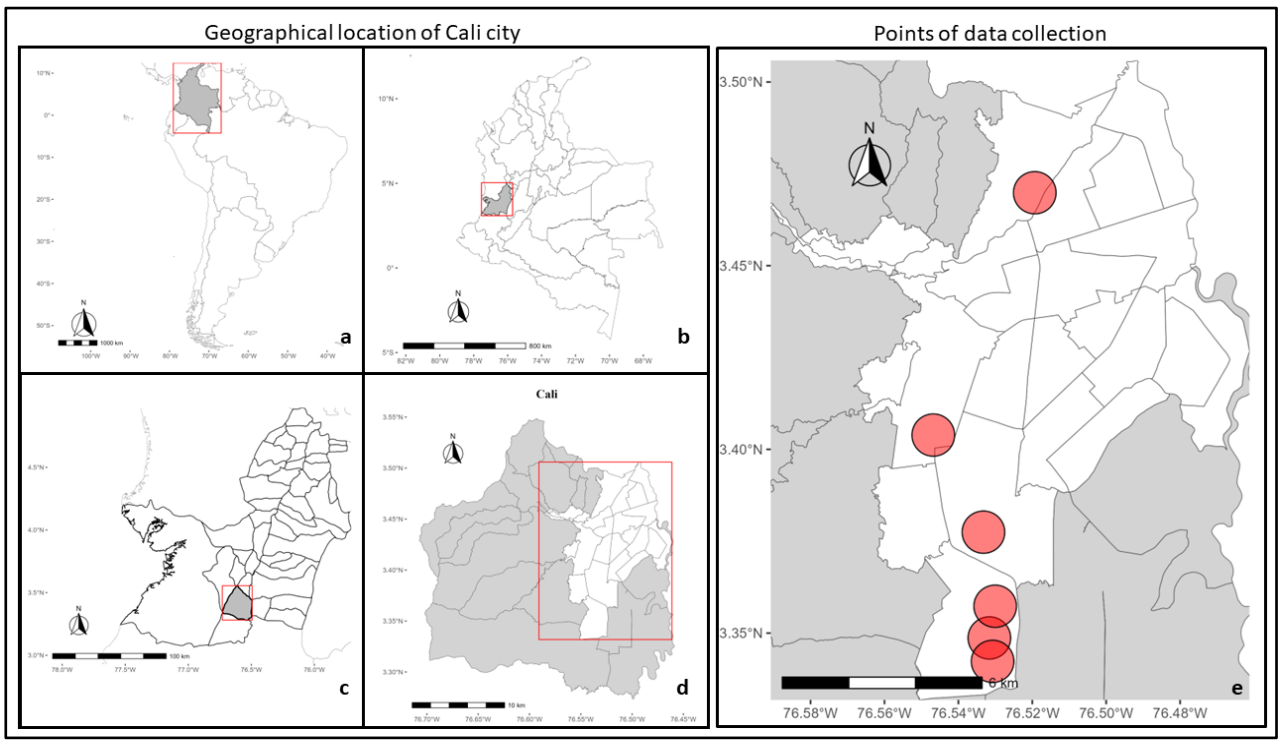

Figure 1. Location of Cali city and (a) Location of Colombia in South America (b) Location of Valle del Cauca State in Colombia (c) Location of Cali city in Valle del Cauca State (d) Cali urban area (e) Districts in which participant organizations are located.

*Descriptive analysis*

In the descriptive analysis, graphical visualization is employed to depict the most relevant results obtained from respondents' answers. Charts are presented that encompass the distribution of transportation modes, the importance of various factors influencing mode choice, as well as travel times and distances. Graphical representations are also provided based on users' socioeconomic stratum and their perception of insecurity. Additionally, differences in mode choice and the importance of influencing factors are explored, segmented by gender. The varying importance of mode choice factors, depending on the mode of transportation used by each user, is visually represented, and a heat map of the city is presented, indicating the average time people from each district take for their daily commutes.

*Comparison between variables*

The comparison between groups, considering the mode of transportation and the socioeconomic level, was conducted using specific tests. For continuous variables, the Mann-Whitney test (MW) and the Kruskal-Wallis test (KW) were applied. In the case of categorical variables, either the chi-square test or Fisher's exact test was employed.

*Cluster analysis*

Initially, the categorical and numerical variables were separated in order to facilitate the manipulation of the data. In total there are 37 numerical and 12 categorical variables. For the implementation of the algorithm, a common condition in most clustering algorithms, which is their definition for numerical data, was faced. To address this, the recoding of categorical variables into binary ones was performed. Suppose, in general, a table with n variables where each has $p_i$ modalities or categories (for $i=1, \ldots, k$). An indicator variable is associated with each modality within each variable or entry. The coding given by $p_i$ corresponds to as many binary variables as modalities the categorical variable has. The total number of modalities is equal $\sum_{i=1}^{k} p_i = p$. For a particular individual, it is encoded with one (1) if the individual has the attribute of the respective modality and with zero (0) in the other modalities of the same variable, assuming that the modalities are mutually exclusive (Díaz Monroy & Morales Rivera, 2009). The optimal number of clusters was two, determined through the Average Silhouette methodology, considering the number of clusters that maximizes the average silhouette coefficient of all observations. The silhouette coefficient ($s_i$) quantifies how good the assignment of an observation is by comparing its similarity to the rest of the observations in its cluster versus those of the other clusters. Its value can range from -1 to 1, with values close to 1 indicating that the observation has been assigned to the correct cluster.

Once the recoding was done and the optimal number of clusters determined, data scaling was performed, and the clustering model was trained using the Sklearn library, with Euclidean distance for calculating distance between instances in the matrix and Ward linkage criterion, which minimizes the variance of merged clusters. Subsequently, the performance of the cluster was analyzed using the F1 score obtained in cross-validation. After further investigation, it was discovered that hierarchical agglomerative clustering for mixed data types is also possible in Python when calculating the distance matrix, as follows (Coding Infinite, 2023).

In the initial development, a function was defined to calculate the distance between two data points with mixed attributes. This process involved calculating dissimilarity between categorical values and the distance between numerical attributes individually. The mixed distance function checks if categorical parameters are null, suggesting the absence of categorical attributes, and calculates the squared Euclidean distance in such cases. If categorical parameters are present, the function segregates categorical and numerical attributes and calculates their distances independently using specific functions for each attribute. The resulting values are then scaled and combined.

After defining the function to calculate the distance between data points, another function is established to calculate the distance matrix for the given dataset, which has attributes with mixed data types. The distance matrix obtained is transferred to the linkage () function of the SciPy Cluster Hierarchy module, thus generating a linkage matrix. With the linkage matrix in hand, the corresponding dendrogram for the dataset containing mixed data types is plotted. Finally, the cumulative clustering model is trained using the Sklearn library, with precalculated distance since it uses a linkage matrix as input, and complete linkage criterion. The "complete" or "maximum" linkage uses the maximum distances between all observations of the two sets. Labels are then applied to different clusters by performing hierarchical agglomerative clustering.

Again, the cluster performance was analyzed using the F1 score obtained in the cross-validation, to compare the results when taking the input data as mixed data.

Regarding the detail of the agglomeration technique, it begins with each data point in its own cluster and iteratively merges the closest pairs of clusters until forming a single cluster containing all data points. The algorithm results in a dendrogram, a tree-shaped diagram showing the hierarchical relationships between clusters. Hierarchical agglomeration can be performed using various linkage criteria to measure the distance between clusters. These criteria include Single Linkage, which measures the distance between the closest pair of points in different clusters, Complete Linkage, which assesses the distance between the farthest pair of points in distinct clusters, and Average Linkage, calculating the average distance between all pairs of points in different clusters. Mathematically, the algorithm can be represented by the following equation:

$$d_{ij} = linkage(C_i, C_j)$$

Where $d_{ij}$ is the distance between clusters $C_i$ y $C_j$, and the linkage is the chosen linkage criterion. The hierarchical clustering algorithm has several advantages, including its ability to handle non-linearly separable data and interpretability through the dendrogram. However, it is computationally expensive for large dataset (Pelekis et al., 2023).

In this study a cluster analysis was conducted on users of different modes of transportation in Cali city. Cluster analysis is employed to examine contingency tables (cross-tabulation). These tables contain information obtained by crossing the modalities of two qualitative variables defined over the same population of *n* individuals. The purpose is to obtain a typology for both rows and columns and establish relationships between these two typologies. In other words, the main objective is to break down the relationship between two variables into a sum (or overlap) of simple and interpretable trends to measure their relative importance. Although this analysis mainly deals with two-dimensional contingency tables, it could also extend to tables with three or more entries. The rows of these tables represent objects or individuals, while the columns represent the modalities of categorical variables. In this context, multiple correspondence analysis is employed, a simple correspondence analysis applied not only to a contingency table but to a complete disjunctive table. The objectives of this analysis are presented based on three families of objects involved. One of the objectives is to create a typology of individuals based on a notion of similarity, where two individuals are closer the greater the number of modalities they have in common. From the perspective of variables, studying the relationship between qualitative variables requires considering the contingency table that crosses their modalities. Finally, examining the set of modalities of variables involves establishing a balance of their similarities (Escofier & Pages, 1992).

It was determined to take a partitioning clustering method to compare with hierarchical clustering in this case study, using the score F1, which ranges between 0 and 1, the higher the value, the more users are grouped into meaningful and easily distinguishable clusters. The k-means method was selected for comparison as it is widely used due to its simplicity and speed, as well as its low computational requirements. This method groups observations into a predefined number of K clusters so that the sum of internal variances of the clusters is minimal (Rodrigo, 2023). These methodologies have also been employed and compared in previous research studies (Pelekis et al., 2023).

Once the methods have been compared, the one with the best performance is selected and its respective dendrogram is made. Furthermore, the characteristics of the clusters generated are visualized and the categories of the clusters are compared using radar plots.

**Results and discussion**

*Descriptive Phase*

The most significant results related to mode of transportation choice are depicted in Figure 2. Initially, it illustrates the modes of transportation used by the respondents, with the car being the predominant mode, followed by public transportation. It also highlights the factors influencing mode of transportation choice. On average, the respondents prefer modes of transportation that offer shorter travel times, safety, and comfort. For this factor assessment, respondents were asked to rate each aspect on a scale of 1 to 5, where 1 signified "not very important" and 5 signified "very important," to gauge how crucial each aspect was in their current mode of transportation selection. The ratings for each factor were averaged to calculate the final score for each factor. Lastly, it demonstrates how the surveyed population is distributed in terms of the duration of their daily commutes and the distances they cover. The majority takes between 21 and 30 minutes for their commutes. It is evident that the car is the preferred choice for individuals in all three distance ranges covered by the respondents: 0 to 10 kilometers, 10 to 20 kilometers, and over 20 kilometers.

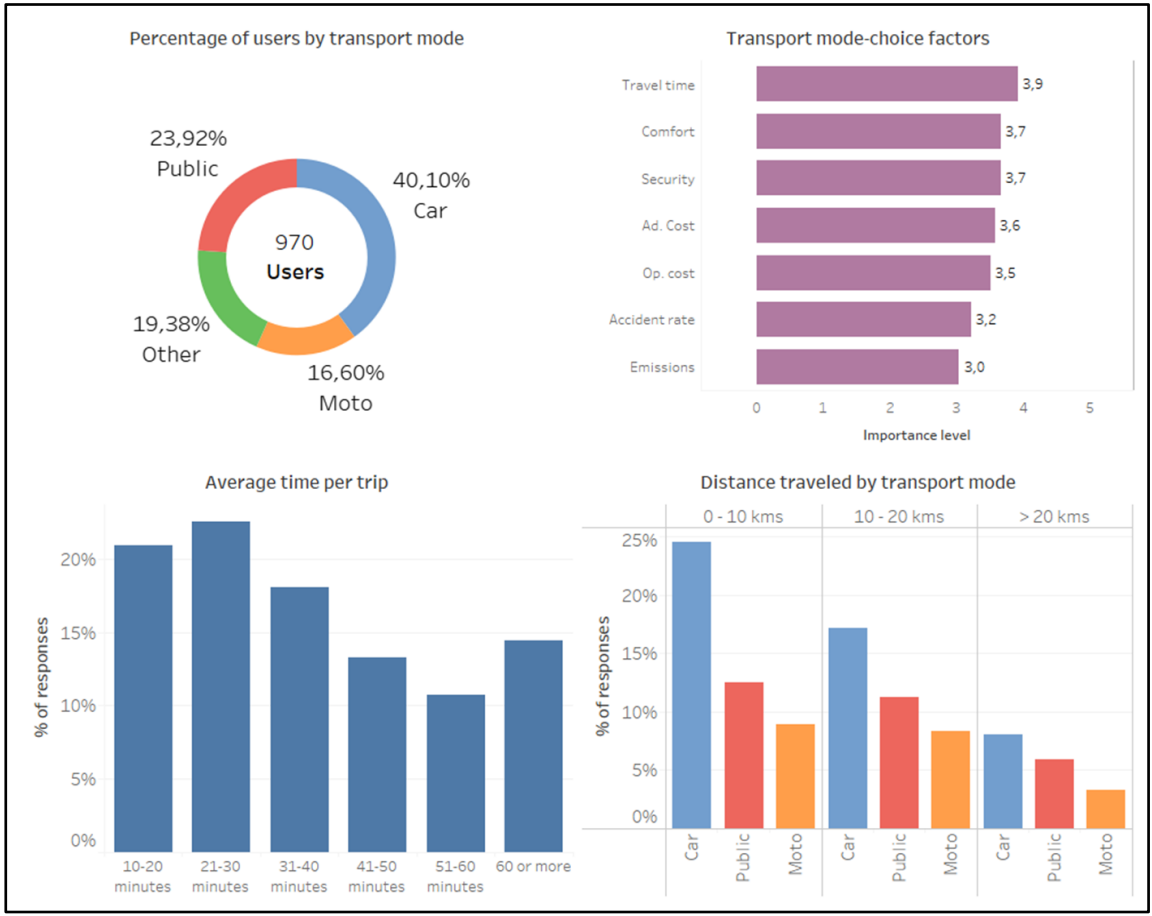

Figure 2. Distribution of survey respondents by transport mode and travel patterns.

A gender-based exploration was also conducted, as shown in Figure 3. It reveals that in all the analyzed modes of transportation, most respondents were men. Regarding the factors influencing mode of transportation choice, it is observed that most female respondents place greater importance on travel time, safety, and comfort compared to men. Most of the surveyed individuals belonged to the middle socioeconomic stratum, followed by the high stratum, with the low stratum ranking third. Lastly, concerning the perception of insecurity, a significant majority of people feel a high level of insecurity when using public transportation.

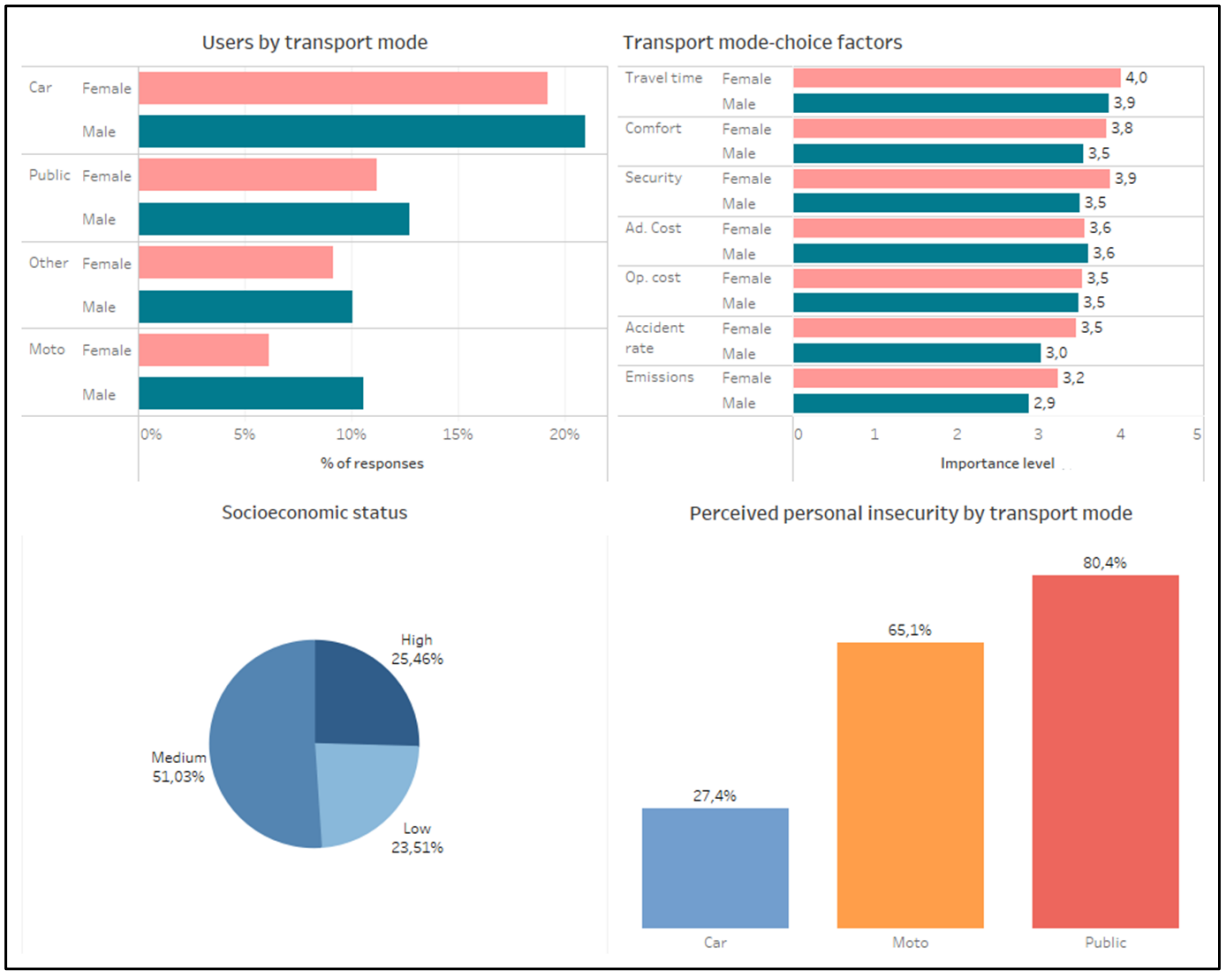

Figure 3. Distribution of survey respondents by demographic characteristics and perceptions.

To compare the importance that users of each mode of transportation assign to each factor, Figure 4 illustrates the findings. It is evident that car users prioritize travel time, comfort, and safety. On the other hand, motorcycle users prioritize costs in addition to travel time. Bus users primarily prioritize costs.

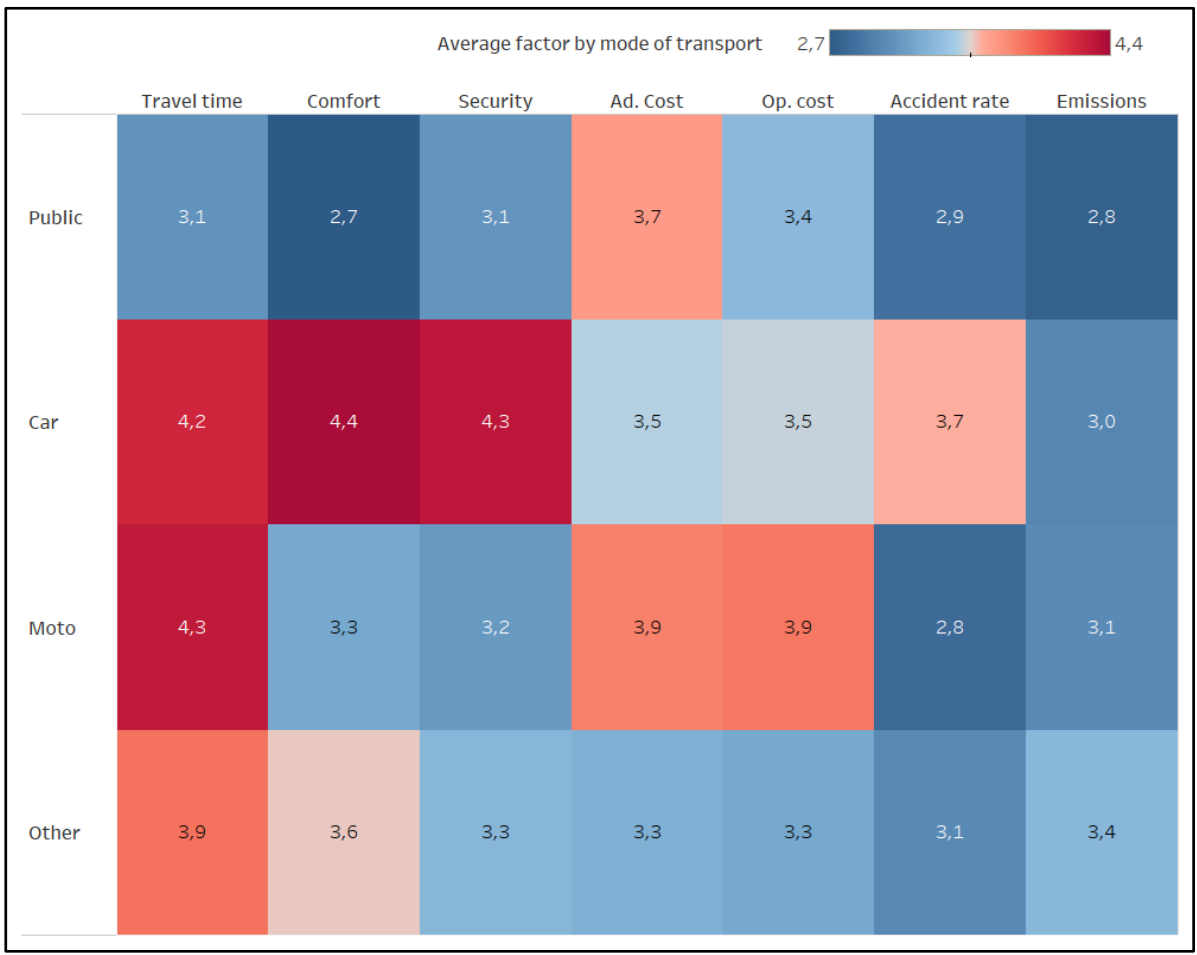

Figure 4. Relation between transport mode and the average perception level for factors influencing the mode-choice.

The georeferencing of information is presented in Figure 5, depicting the average time it takes individuals from each district to commute in their daily journeys. It is important to note that, in general, the georeferencing does not account for users residing in nearby municipalities and users who commute to a different area daily. It is observed that people residing in the northern part of the city take the longest time for these commutes, while those residing further south tend to have shorter travel times.

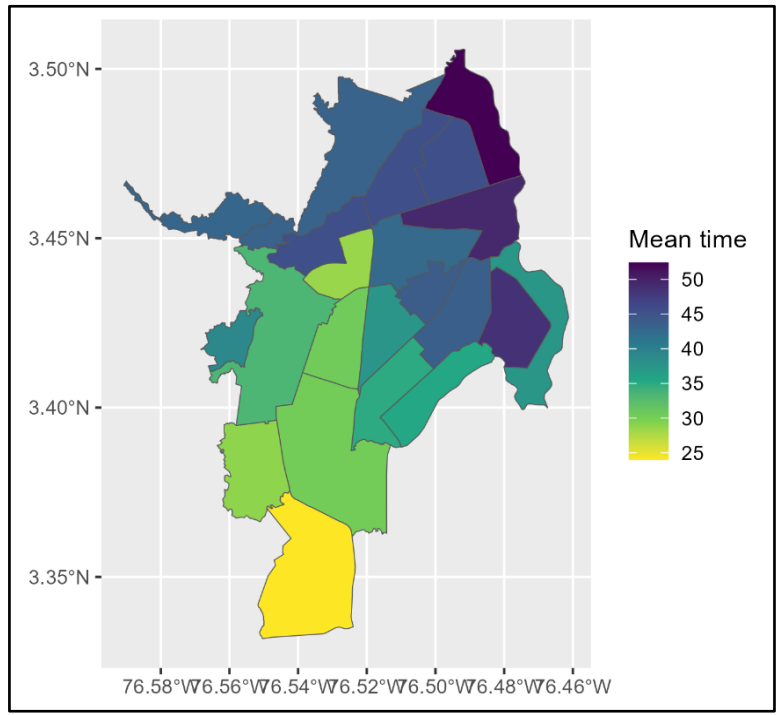

Figure 5. Average travel time per trip by origin districts in Cali city.

Additionally, a public online digital dashboard was created, which allows the application of dynamic filters so that the graphs present a greater amount of information depending on the interests of the person who consults it.

*Comparison between variables*

In terms of participation, the following cross-tabulated tables are considered: first, Table 2 presents the proportion of users differentiated by gender and age groups; Table 3, the proportion differentiated by the main mode of urban transportation and age; and Table 4, the proportion differentiated by the main mode of urban transportation and socioeconomic stratum. Please note that, in general, the following recategorization is applied to the socioeconomic stratum: users in low-stratum households correspond to users in very low (stratum 1) or low (stratum 2) households; users in middle-stratum households correspond to users in lower middle (stratum 3) or middle (stratum 4) households; and users in high-stratum households correspond to users in upper middle (stratum 5) or high (stratum 6) households.

Table 2. Proportion of participants differentiated by gender and age groups.

| Age | Female | Male |
|---|---|---|
| 15 - 29 years | 26,97% | 34,54% |
| 30 - 49 years | 14,42% | 12,34% |
| 50 years or older | 4,25% | 7,47% |

Table 3. Proportion differentiated by the main mode of urban transportation and age.

| Age | Active | App | Car | Motorcycle | Taxi | Informal | Public |
|---|---|---|---|---|---|---|---|
| 15 - 29 years | 6,33% | 4,56% | 18,98% | 9,34% | 0,00% | 2,18% | 20,12% |
| 30 - 49 years | 2,49% | 1,76% | 13,28% | 5,71% | 0,10% | 0,21% | 3,22% |
| 50 years or older | 0,83% | 0,21% | 7,88% | 1,66% | 0,41% | 0,10% | 0,62% |

Table 4. Proportion differentiated by the main mode of urban transportation and socioeconomic stratum.

| Mode | High | Low | Medium |
|---|---|---|---|
| Active | 2,28% | 2,07% | 5,29% |
| App | 2,28% | 0,52% | 3,73% |
| Car | 17,74% | 2,39% | 20,02% |
| Motorcycle | 1,04% | 6,85% | 8,82% |
| Taxi | 0,10% | 0,21% | 0,21% |
| Informal | 0,10% | 1,35% | 1,04% |
| Public | 1,76% | 10,17% | 12,03% |

In general, the following perception scales are considered as continuous variables: one, the importance of the acquisition cost in the choice of the main mode of urban transportation (acq_cost); two, the importance of operating cost (op_cost); three, the importance of safety in terms of accidents (accidents); four, the importance of safety in terms of crime (crime); five, the importance of material comfort (comfort); six, the importance assigned to the level of atmospheric pollutant emissions; and seven, the importance of travel time (time). Note that, according to Shapiro-Wilk normality test, the considered continuous variables do not follow a normal distribution (see Table 8 attached).

Considering the results of the Shapiro-Wilk normality test, Table 5 presents the median and interquartile range, along with the mean and standard deviation associated with each one. In the case of categorical variables, the count and proportion are reported. Based on the mode of transportation and socioeconomic level, the comparison between groups for continuous variables is conducted using the Mann-Whitney (MW) test and the Kruskal-Wallis (KW) test for k > 2 populations. For categorical variables, the chi-square test or Fisher's exact test is employed. For a variable with k categories, the non-parametric Kruskal-Wallis test is implemented to determine if the k populations are identical (see Table 9 attached).

Table 5. Variables by mode of transport.

| Variable | Measurement | All | Car | M/cycle | Bus | Other | p-value |
|----------|-------------|-----|-----|---------|-----|-------|---------|
| accidents (LTS) | Q2 (IQR) | 3 (2, 4) | 4 (3, 5) | 3 (2, 3) | 3 (2, 4) | 3 (2, 4) | |
| | Mean (sd) | 3.22(1.32) | 3.65(1.22) | 2.84(1.12) | 2.88(1.35) | 3.09(1.41) | 0*** |
| comfort (LTS) | Q2 (IQR) | 4 (3, 5) | 5 (4, 5) | 3 (3, 4) | 3 (1, 4) | 4 (3, 5) | |
| | Mean (sd) | 3.67(1.36) | 4.42(0.95) | 3.25(1.19) | 2.74(1.38) | 3.61(1.34) | 0*** |
| acq_cost (LTS) | Q2 (IQR) | 4 (3, 5) | 4 (3, 4) | 4 (3, 5) | 4 (3, 5) | 3 (2.5, 4.5) | 0*** |
| op_cost (LTS) | Mean (sd) | 3.58(1.23) | 3.52(1.15) | 3.85(1.07) | 3.73(1.3) | 3.3(1.38) | |
| | Q2 (IQR) | 4 (3, 4) | 4 (3, 4) | 4 (3, 5) | 4 (2, 5) | 3 (2, 4) | |
| crime (LTS) | Mean (sd) | 3.5(1.26) | 3.54(1.02) | 3.89(1.06) | 3.37(1.54) | 3.23(1.42) | 0*** |
| | Q2 (IQR) | 4 (3, 5) | 5 (4, 5) | 3 (2, 4) | 3 (2, 4) | 3 (3, 4) | |
| emissions (LTS) | Mean (sd) | 3.66(1.29) | 4.34(0.99) | 3.16(1.18) | 3.14(1.28) | 3.32(1.3) | 0*** |
| | Q2 (IQR) | 3 (2, 4) | 3 (2, 4) | 3 (2, 4) | 3 (1, 4) | 3 (2.5, 5) | |
| satisfaction (LTS) | Mean (sd) | 3.04(1.25) | 3.03(1.13) | 3.06(1.16) | 2.78(1.34) | 3.35(1.36) | 0*** |
| | Q2 (IQR) | 4 (3, 5) | 5 (4, 5) | 4 (4, 5) | 2 (1, 3) | 4 (3, 5) | |
| travel.time (LTS) | Mean (sd) | 3.89(1.26) | 4.56(0.65) | 4.35(0.78) | 2.45(1.23) | 3.9(1.09) | 0*** |
| | Q2 (IQR) | 4 (3, 5) | 5 (4, 5) | 5 (4, 5) | 3 (2, 4) | 4 (3, 5) | |
| | Mean (sd) | 3.92(1.26) | 4.24(0.98) | 4.32(1.09) | 3.11(1.4) | 3.9(1.3) | 0*** |
| Gender, n (%) | | | | | | | |
| Female | | 440 (45.64%) | 185 (47.8%) | 59 (36.65%) | 108 (46.75%) | 88 (47.57%) | 0*** |
| Male | | 524 (54.36%) | 202 (52.2%) | 102 (63.35%) | 123 (53.25%) | 97 (52.43%) | |
| SES, n (%) | | | | | | | |
| High | | 244 (25.31%) | 171 (44.19%) | 10 (6.21%) | 17 (7.36%) | 46 (24.86%) | 0,33 |
| Low | | 227 (23.55%) | 23 (5.94%) | 66 (40.99%) | 98 (42.42%) | 40 (21.62%) | |
| Medium | | 493 (51.14%) | 193 (49.87%) | 85 (52.8%) | 116 (50.22%) | 99 (53.51%) | |
| Occupation, n (%) | | | | | | | |
| Employee | | 405 (42.01%) | 201 (51.94%) | 92 (57.14%) | 44 (19.05%) | 68 (36.76%) | 0,16 |
| Student | | 544 (56.43%) | 177 (45.74%) | 67 (41.61%) | 185 (80.09%) | 115 (62.16%) | |
| Other | | 15 (1.56%) | 9 (2.33%) | 2 (1.24%) | 2 (0.87%) | 2 (1.08%) | |
| Age, n (%) | | | | | | | |
| 15 - 29 years | | 593 (61.51%) | 183 (47.29%) | 90 (55.9%) | 194 (83.98%) | 126 (68.11%) | 0*** |
| 30 - 49 years | | 258 (26.76%) | 128 (33.07%) | 55 (34.16%) | 31 (13.42%) | 44 (23.78%) | |
| 50 years or older | | 113 (11.72%) | 76 (19.64%) | 16 (9.94%) | 6 (2.6%) | 15 (8.11%) | |

LTS = Likert-type scale. SES = Socioeconomic-Status. p-value = p-value for Kruskal-Wallis, chi-2 or Fisher's exact test, as appropiate. (***p-value <0,01, **p-value <0,05, *p-value <0,1)

Note that based on the results of the Kruskal-Wallis non-parametric test for k categories, the implementation of the Mann-Whitney non-parametric test for pairwise categories is proposed. In Figure 6, the results are observed differentiated by pairs of socioeconomic status (SS).

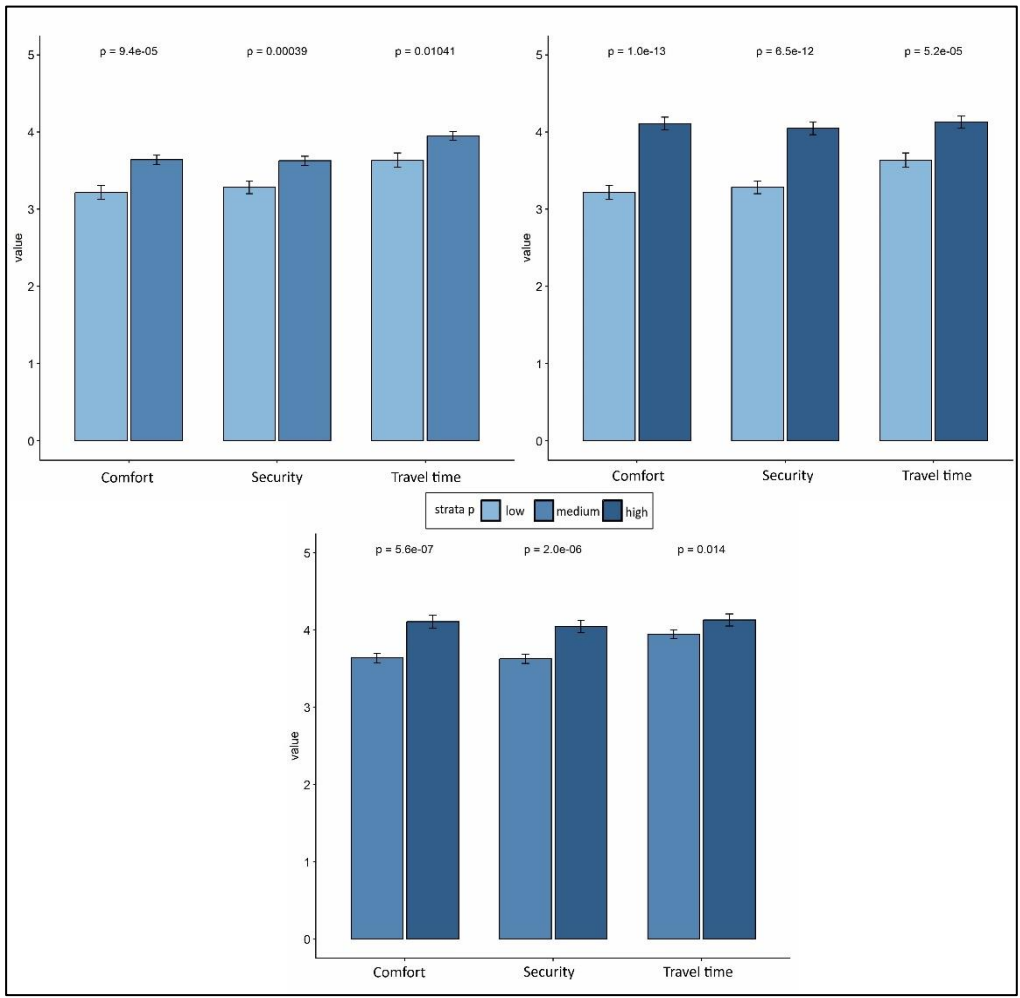

Figure 6. Non-parametric Mann-Whitney U test (Socioeconomic status).

Lastly, in a similar vein, to determine the independence between categorical variables, Table 6 presents the results of implementing the χ2 independence test. Additionally, the Fisher exact test is performed to ensure that both tests lead to identical conclusions.

Table 6. Chi-square (χ2) test of independence.

| $Z_1$ | $Z_2$ | χ2 | df | p-value |
|---|---|---|---|---|
| gender | SS | 2,20425 | 2 | 0,33217 |
| gender | mode | 19,767 | 7 | 0*** |
| gender | occupation | 3,6887 | 2 | 0,15813 |
| SS | mode | 239,66 | 14 | 0*** |
| SS | occupation | 34,1562 | 4 | 0*** |
| mode | occupation | 103,428 | 14 | 0*** |
| mode | age | 130,4953 | 14 | 0*** |
| **Null hypothesis:** the two variables are independent. *** $p-value < 0,01$, ** $p-value < 0,05$, * $p-value < 0,1$. Fisher's exact test leads to the same conclusions | | | | |

In Figure 7, two grouped variables are considered: first, the average daily travel time by urban transportation users (Time); and second, the average daily distance traveled from the participant's home to their frequent destination (Distance). The first variable consists of six classes with a width of 9 minutes (namely, from 11 to 20 minutes, from 21 to 30 minutes, from 31 to 40 minutes, from 41 to 50 minutes, from 51 to 60 minutes, and from 61 to 70 minutes); the second variable has five classes with a width of 4 kilometers (namely, from 1 to 5 kilometers, from 6 to 10 kilometers, from 11 to 15 kilometers, from 16 to 20 kilometers, and from 21 to 30 kilometers). For simplicity, for each grouped variable, the mean with its standard error by mode of transport is reported.

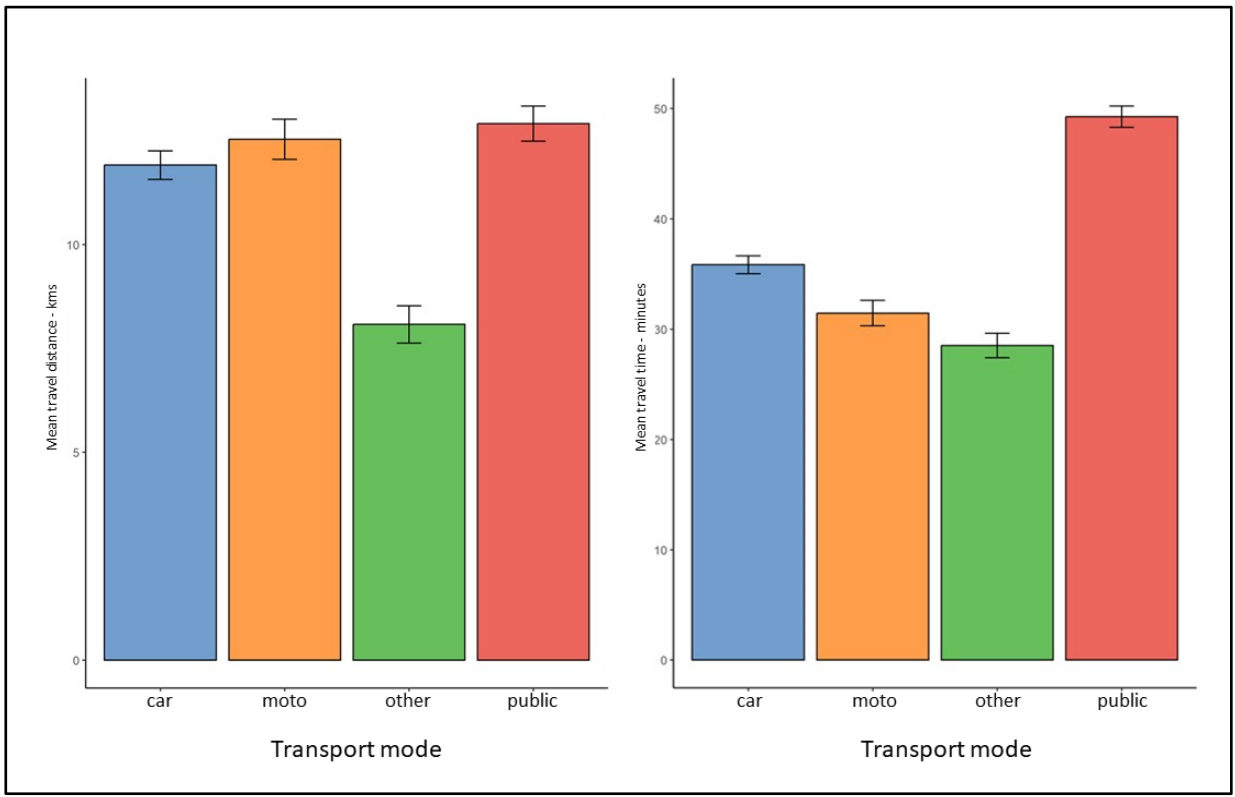

Figure 7. Grouped variables by transport mode.

*Cluster analysis*

The agglomerative technique using a mixed data set obtained an F1 score of 0.9518, a considerably better performance than the k-means method, indicating that users are grouped into meaningful and easily distinguishable clusters. Then, an analysis is made of the results of the groups generated by this methodology. A dendrogram, a tree-shaped diagram representing the clustering process and displaying the hierarchical links between groups, was generated to showcase the results of hierarchical clustering. This was performed with Euclidean distance for calculating the distance between instances in the matrix and Ward linkage criterion (Figure 8).

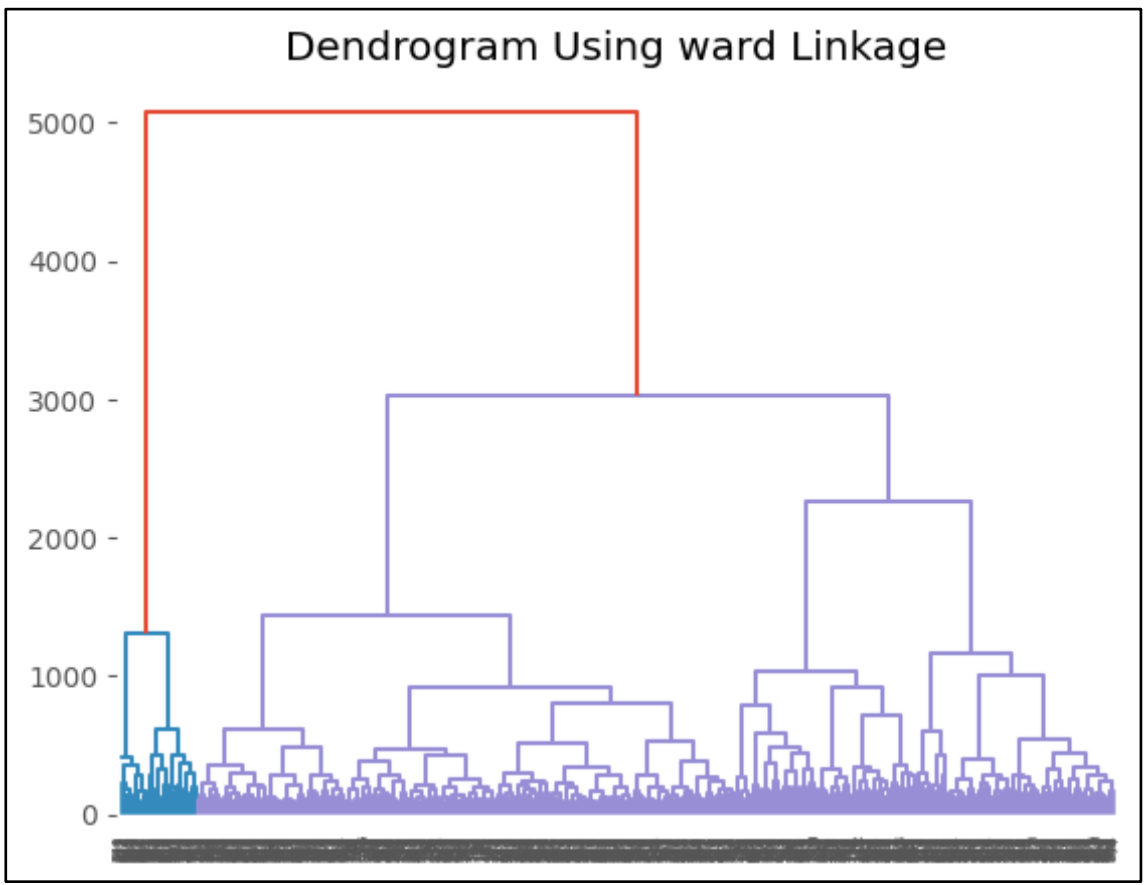

Figure 8. Cluster dendrogram.

Table 7 summarizes the sociodemographic characteristics of the 970 users included in the study. It is relevant to note that, in all cases, the proportions between variables in each category are similar for each grouping. However, in terms of occupation, Group 2 is characterized by not including any unemployed users. The majority of participants fall within the age range of 15 to 29 years, predominantly composed of men who primarily use the car as a means of transportation and work as employees.

Table 7. Characteristics by cluster.

| Cluster | 1 (N=803) | 2 (N=167) | Total (N=970) |

| Age | | | |
|---|---|---|---|
| 15 - 29 years | 470 | 129 | 599 |
| 30 - 49 years | 236 | 22 | 258 |
| 50 years or older | 97 | 16 | 113 |
| Sex | | | |
| Female | 367 | 73 | 440 |
| Male | 432 | 92 | 524 |
| SS | | | |
| Low | 184 | 44 | 228 |
| Medium | 418 | 77 | 495 |
| High | 201 | 46 | 247 |
| Mode of transport | | | |
| Car | 318 | 71 | 389 |
| Motorcycle | 141 | 20 | 161 |
| Public transport | 190 | 42 | 232 |
| Other | 154 | 34 | 188 |
| Occupation | | | |
| Unemployed | 3 | 0 | 3 |
| Employed | 365 | 40 | 405 |
| Student | 424 | 125 | 549 |
| Other | 11 | 2 | 13 |

When examining the factors influencing the choice of transportation mode, it is observed that, on average, on a scale from 1 to 5 (where 1 is not important at all and 5 is very important), users in Group 1 assign greater importance to each factor compared to those in group 2 (Figure 9).

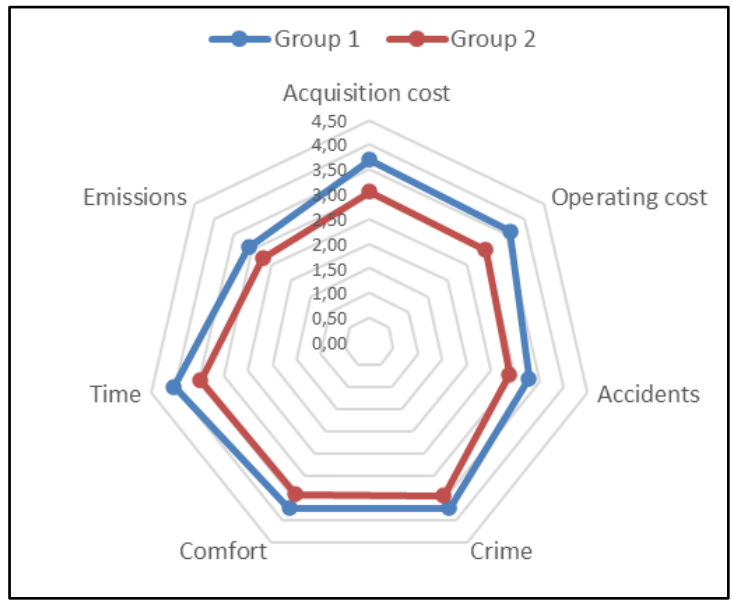

Figure 9. Factors influencing the choice of transportation mode.

In a more detailed analysis, when exploring the average perception of the attributes of each mode of transportation, where 1 means totally disagree and 5 means totally agree, it is found that people in Group 1, on average, give higher scores to each attribute. These attributes range from the economic costs of the transportation mode to accident rates, insecurity, comfort, travel time, and emissions (Figure 10).

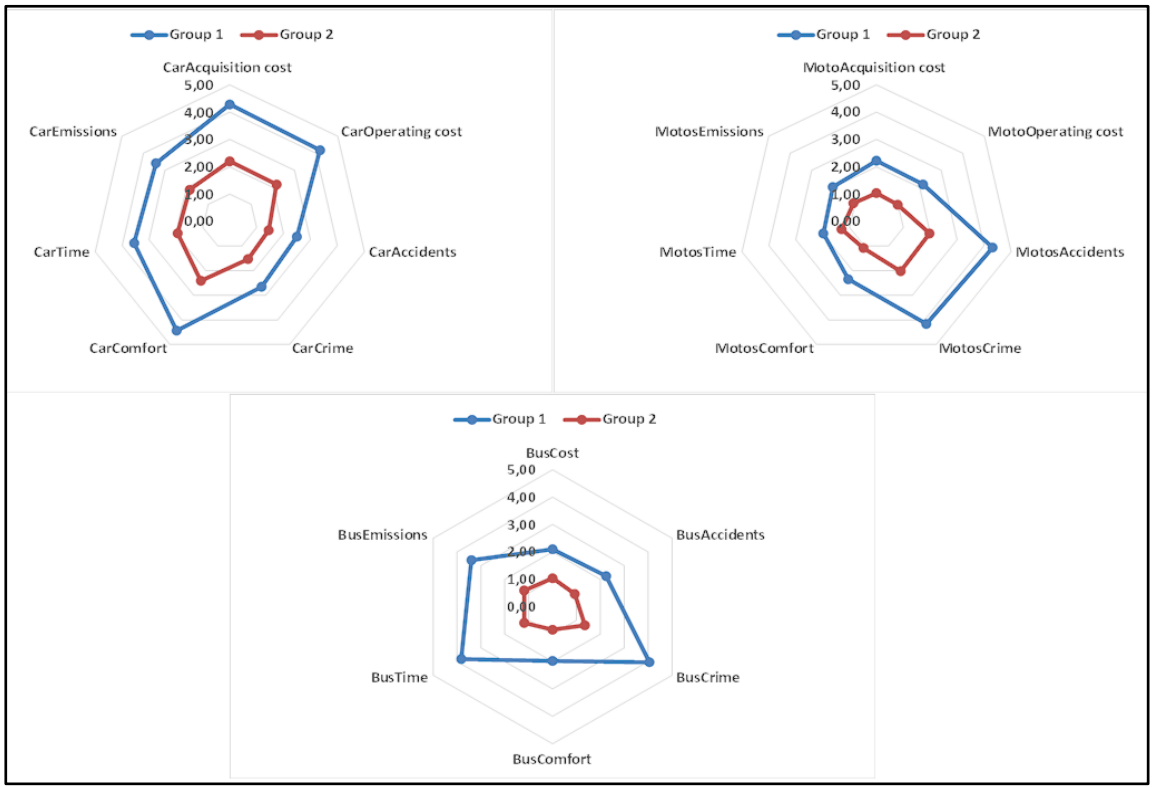

Figure 10. Average perception of the attributes of different transportation modes (car, motorcycle, bus).

To explore the differences among the generated groups, detailed analyses were conducted on the factors influencing the choice of transportation mode and the average perception of the attributes of each mode. This exploration considered variables such as gender, socioeconomic stratum, transportation mode, and the occupation of the respondents (see attached Figures 11-26). The results highlight the following:

Gender: It is observed that both men and women in both groups exhibit similar behaviors in choosing transportation modes and in their perceptions of the three modes. Overall, Group 2 tends to assign lower values to all questions compared to Group 1.

Socioeconomic Stratum: In the choice of transportation mode, it stands out that, in Group 2 with a high socioeconomic stratum, a higher value is assigned to safety compared to Group 1 of the same stratum. Regarding the perception of car attributes, Group 1 maintains a consistent behavior, while in Group 2, as the stratum increases, the scores assigned to the attributes also increase. In the perception of motorcycle attributes, safety and accident rates stand out with high values, and as the stratum increases, the value assigned to the cost of acquisition decreases. For the perception of bus attributes, as the stratum increases, the scores for comfort, cost, and emissions decrease.

Mode of Transportation: In both groups, motorcycle users value time and costs more, car users value time, comfort, and safety more, bus users value costs more, and users of other modes value time more. In the perception of car attributes, car users in Group 2 assign values similar to those in Group 1 , while for other modes, the values assigned by Group 2 are consistently lower. For motorcycle attributes, Group 2 users assign a higher value than Group 1 users for travel time.

Occupation: In the choice of transportation mode, Group 2 users with the occupation "others" assign a high value to time, comfort, and safety compared to those in Group 1. However, they assign a very low value to cost. In the perception of car attributes, Group 2 users with the occupation "others" assign a very low value to accident rates. Regarding the perception of motorcycle attributes, Group 2 users with the occupation "others" assign a value of zero to acquisition and costs, and a value very close to zero to comfort. Again, Group 2 users with the occupation "others" assign a value of zero to travel time, emissions, and accident rates in the perception of bus attributes.

**Conclusions**

The obtained results provide a comprehensive insight into the distinctive characteristics of urban transport users in developing country contexts, with a particular focus on Cali city. The selection of Cali as a case study proved crucial in understanding the decision-making process in an environment where the pivotal role of motorcycles for commuters in low and middle-income classes is prominent.

The information gathered through surveys revealed significant patterns in the choice of urban transport modes, highlighting the decisive influence of socioeconomic stratification in these decisions. Meaningful relationships were identified between transportation choices and variables such as socioeconomic status and age. Furthermore, analyses indicate that in Cali, the choice of transportation mode is strongly influenced by considerations such as travel time, safety, and comfort, revealing notable variations across genders, socioeconomic status, and preferred modes of transport. The geographical distribution of daily travel times underscores specific areas of the city facing significant mobility challenges.

Regarding the employed methodology, the combination of surveys, statistical analysis, and clustering techniques provided a profound understanding of the behavioral patterns of transport users in Cali. Non-parametric statistical tests, such as the Kruskal-Wallis test and cluster analysis, allowed for a rigorous evaluation of differences between groups, ensuring the robustness of the findings. Cluster analysis revealed significant patterns in the behaviors of urban transport users in Cali. By grouping individuals with similar profiles in terms of factors such as transport preferences and perceptions, distinctive segments of the population were identified. These segments provide a deeper understanding of the specific preferences and needs of different user groups in the context of urban transport.

It is crucial to note that, while the presented results offer a valuable perspective on the characteristics of transport users in Cali, the intrinsic difficulty of obtaining a fully representative sample from all areas of the city is acknowledged. The geographical complexity and diversity of urban contexts in Cali pose a significant challenge in encompassing all experiences and perceptions of transportation. Despite these challenges, the collected findings provide substantial insights contributing to the overall understanding of mobility patterns in the city. By better understanding the factors influencing transport decisions, a crucial foundation is laid for the development of policies and urban strategies that effectively address mobility challenges in developing cities, thereby meeting the objectives set at the beginning of this research.

**Acknowledgements**

We would like to thank CASOS at Carnegie Mellon University and CBIE at Arizona State University for providing support to the PI of the project with the wider research on which this paper is based.

**Disclosure Statement**

The authors report there are no competing interests to declare.

**Funding**

This work was supported by the *"Fondo de Ciencia, Tecnología e Innovación del Sistema General de Regalías"* under grant number BPIN 2022000100068.

**Appendices**

Table 8. Shapiro-Wilk normality test for continuous variables.

| Variable | Kurtosis | Skewness | W | p_value |
|---|---|---|---|---|
| acq_cost | 2,59 | -0,68 | 0,87 | 0*** |
| op_cost | 2,43 | -0,62 | 0,87 | 0*** |
| accidents | 1,97 | -0,19 | 0,9 | 0*** |
| crime | 2,35 | -0,63 | 0,86 | 0*** |
| comfort | 2,25 | -0,69 | 0,83 | 0*** |
| travel.time | 2,95 | -1,01 | 0,79 | 0*** |
| emissions | 2,15 | -0,07 | 0,91 | 0*** |
| satisfaction | 3,08 | -1,05 | 0,8 | 0*** |

> **Null hypothesis**: the sample comes from a Gaussian distribution.
>
> ∗∗∗p − value < 0,01, ∗∗p − value < 0,05, ∗p − value < 0,1.

Table 9. Non-parametric Kruskal-Wallis test.

|  | Gender |  | SS |  | Mode |  | Occupation |  |
| --- | --- | --- | --- | --- | --- | --- | --- | --- |
|  | $\chi^2$ (df) | p-value | $\chi^2$ (df) | p-value | $\chi^2$ (df) | p-value | $\chi^2$ (df) | p-value |
| accidents | 24.6 (1) | 0*** | 11.7 (2) | 0** | 77.1 (7) | 0*** | 19.1 (2) | 0*** |
| comfort | 9.7 (1) | 0*** | 62.6 (2) | 0*** | 282.8 (7) | 0*** | 32 (2) | 0*** |
| acq_cost | 0.7 (1) | 0,39 | 3.9 (2) | 0,14 | 26.7 (7) | 0*** | 2.1 (2) | 0,35 |
| op_cost | 0.1 (1) | 0,73 | 4.8 (2) | 0,09 | 25.6 (7) | 0*** | 6.3 (2) | 0** |
| crime | 19.5 (1) | 0*** | 51.8 (2) | 0*** | 238.4 (7) | 0*** | 11.5 (2) | 0*** |
| emissions | 18.7 (1) | 0*** | 1.3 (2) | 0,52 | 39.6 (7) | 0*** | 22 (2) | 0*** |
| satisfaction | 3.7 (1) | 0* | 68.6 (2) | 0*** | 398.9 (7) | 0*** | 50.6 (2) | 0*** |
| travel.time | 2.4 (1) | 0,12 | 17.3 (2) | 0*** | 127.9 (7) | 0*** | 30.5 (2) | 0*** |

> **Null hypothesis**: The k-populations have identical distributions. df = degrees of freedom.
>
> ∗∗∗p − value < 0,01, ∗∗p − value < 0,05, ∗p − value < 0,1.

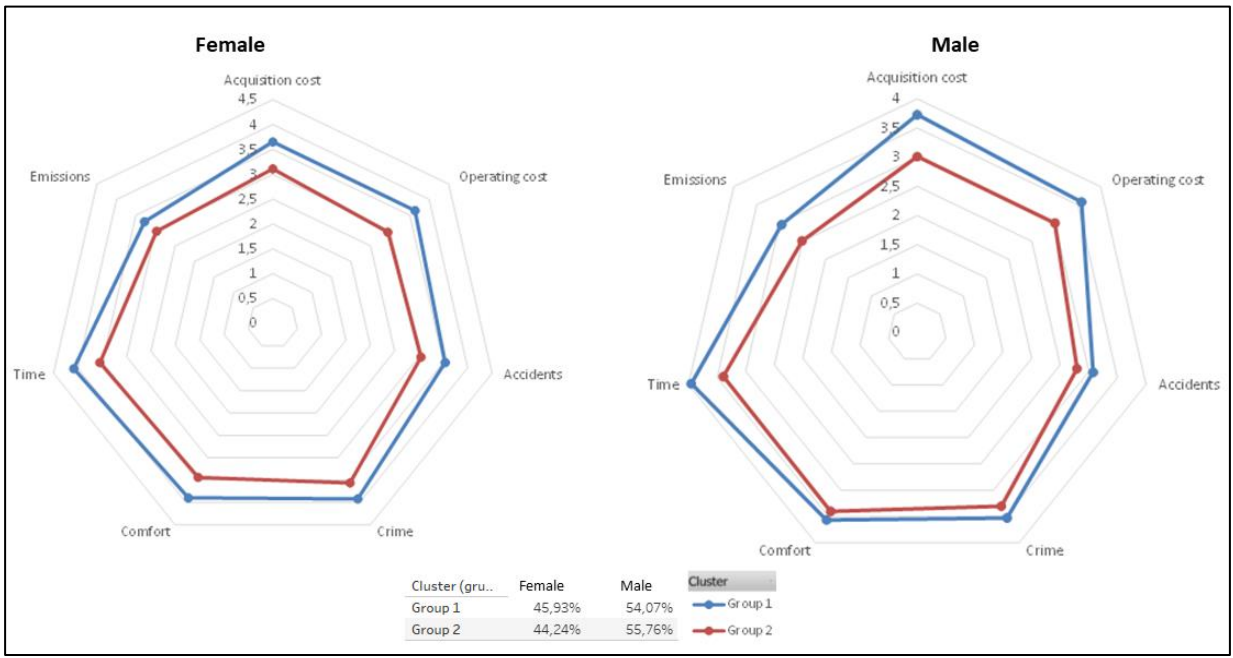

Figure 11. Factors influencing the choice of transportation mode differentiated by gender.

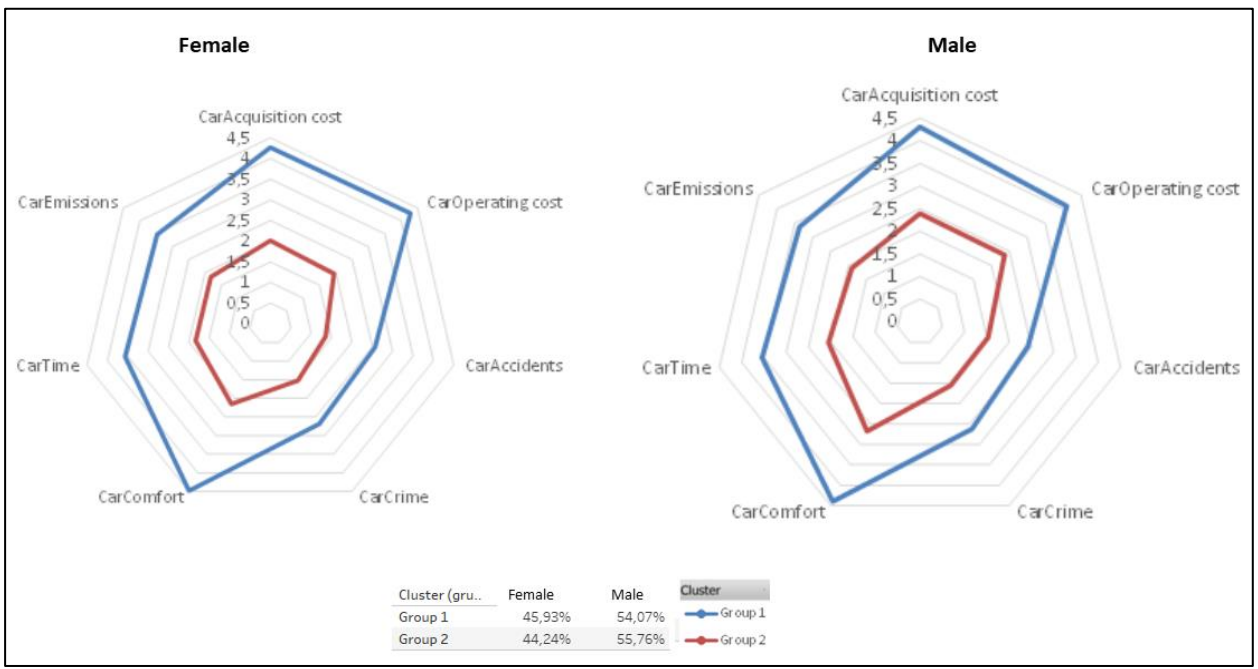

Figure 12. Average perception of car attributes differentiated by gender.

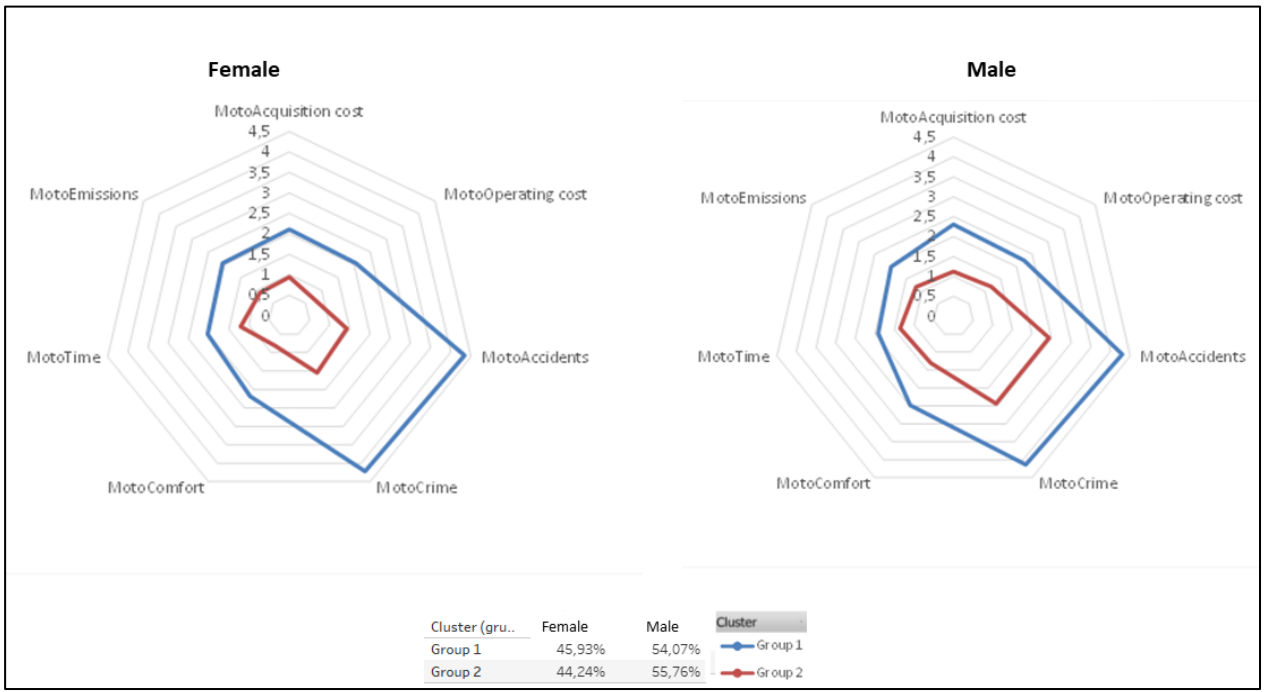

Figure 13. Average perception of motorcycle attributes differentiated by gender.

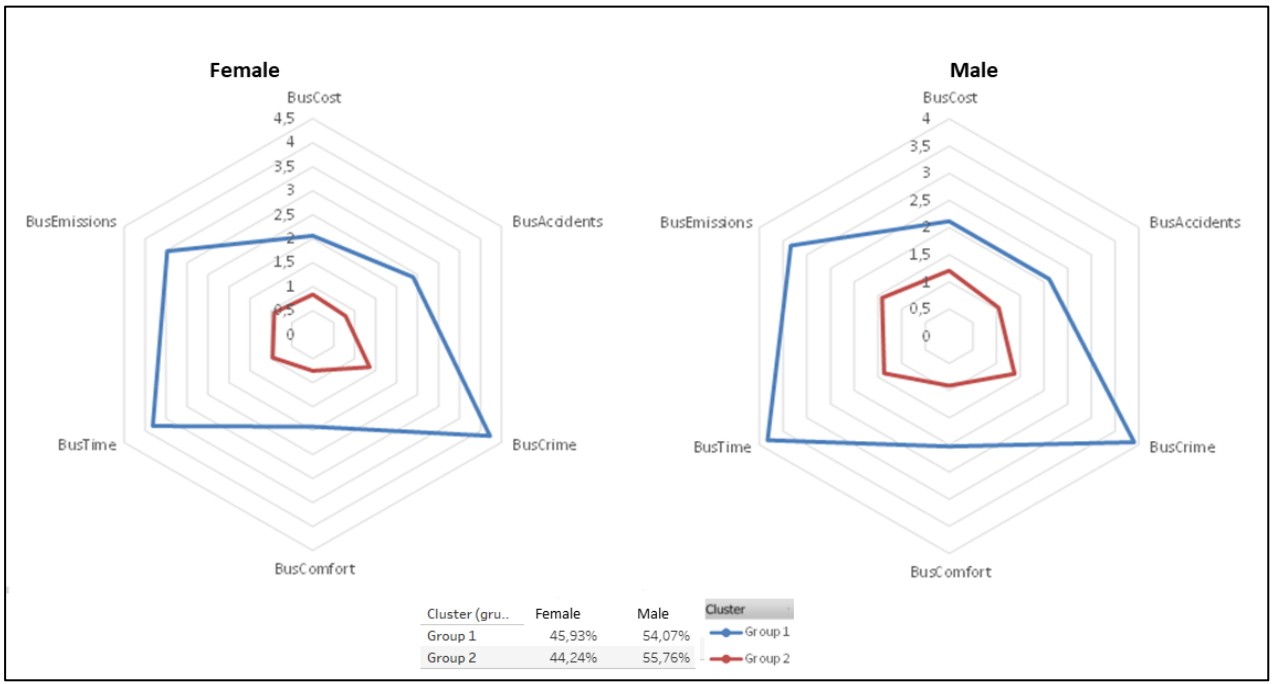

Figure 14. Average perception of public transport attributes differentiated by gender.

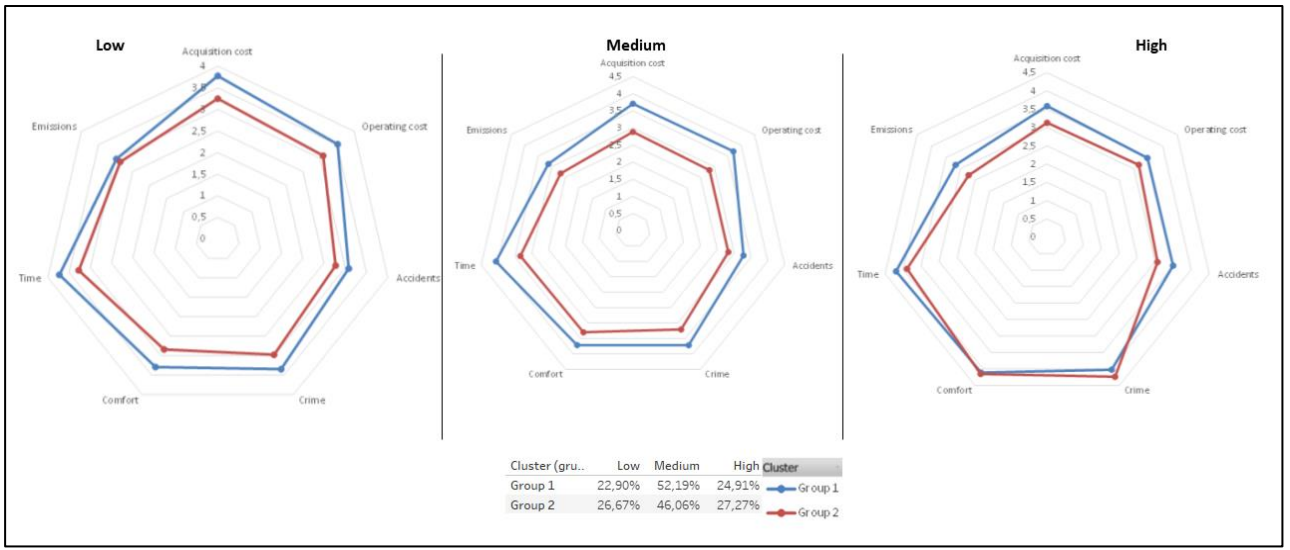

Figure 15. Factors influencing the choice of transportation mode differentiated by socioeconomic status.

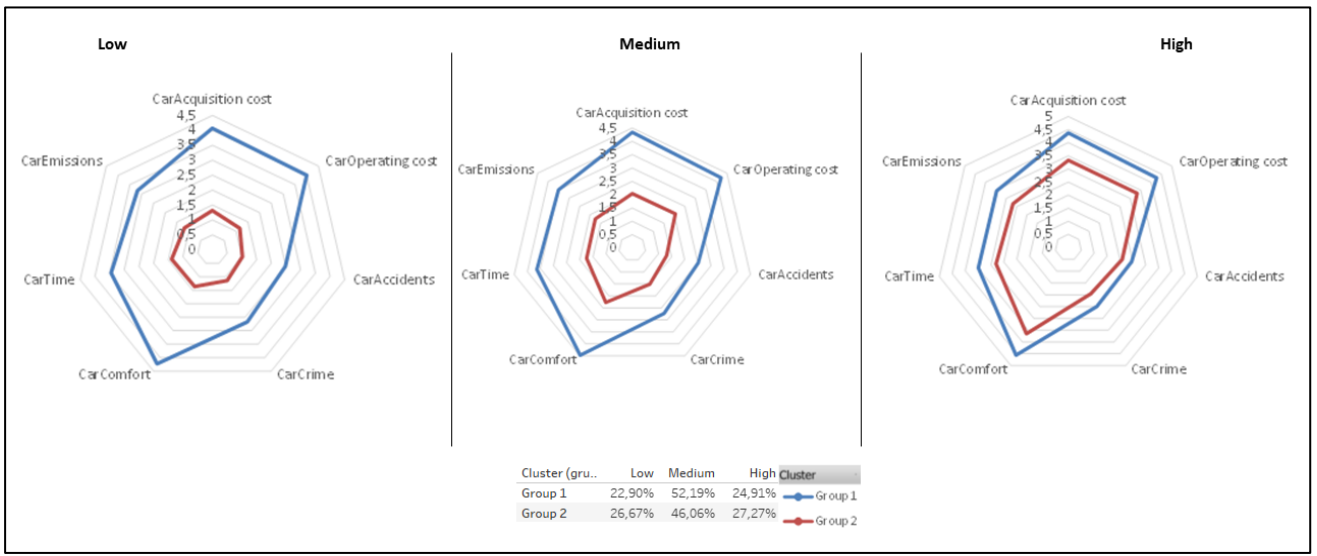

Figure 16. Average perception of car attributes differentiated by socioeconomic status.

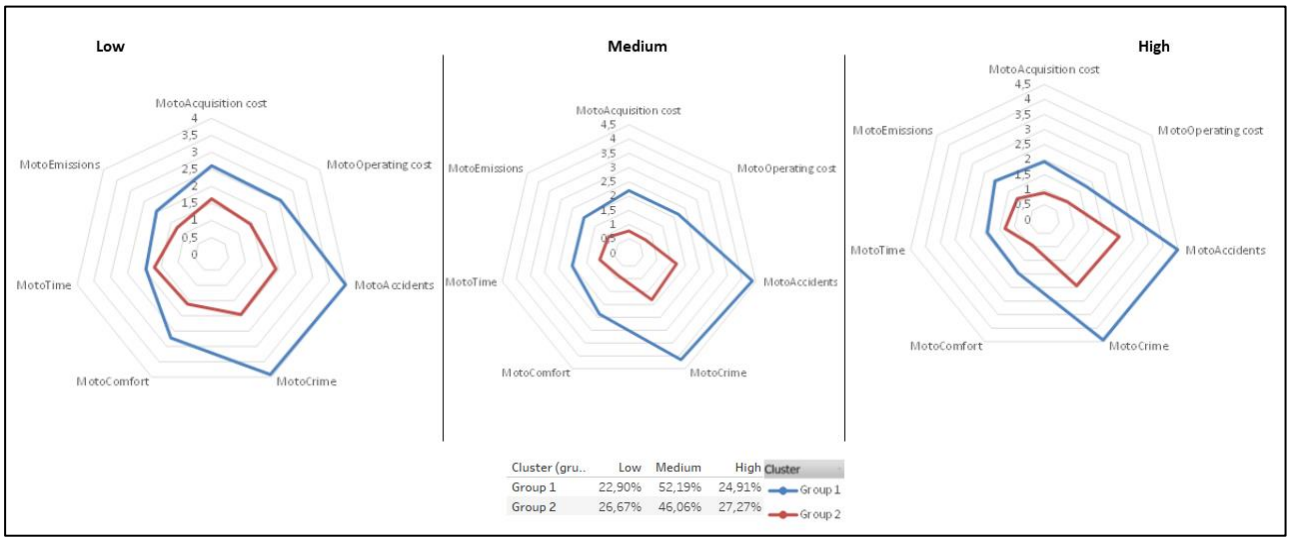

Figure 17. Average perception of motorcycle attributes differentiated by socioeconomic status.

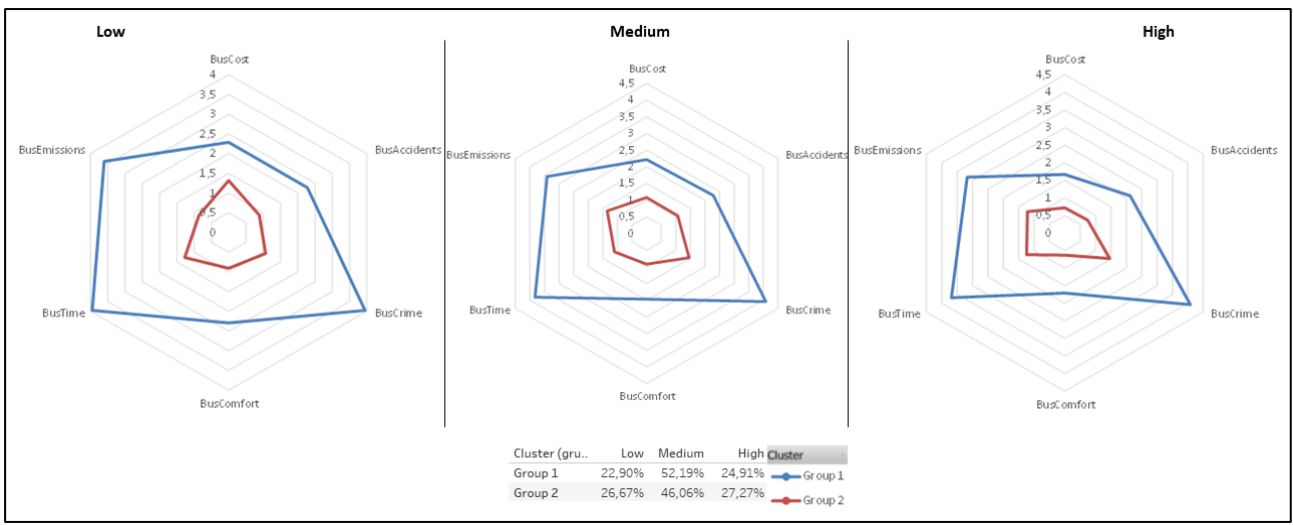

Figure 18. Average perception of public transport attributes differentiated by socioeconomic status.

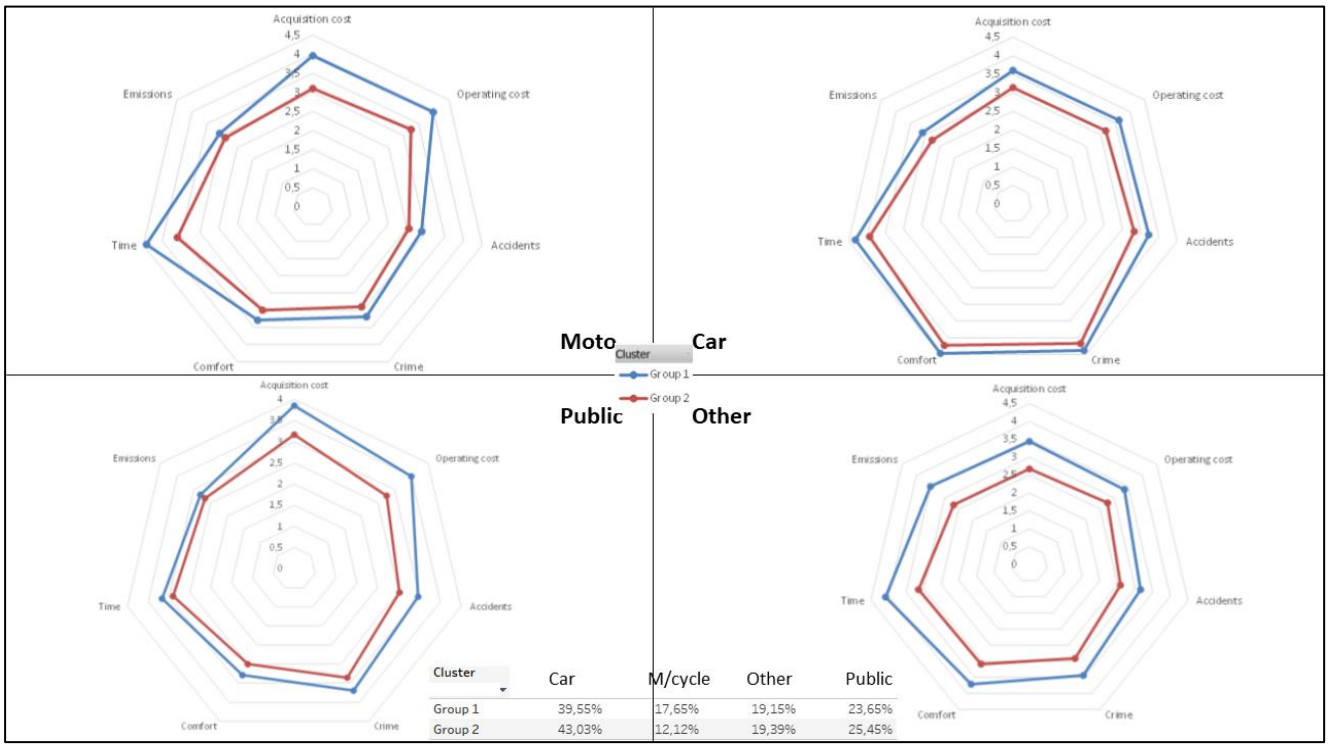

Figure 19. Factors influencing the choice of transportation mode differentiated by users' actual mode of transport.

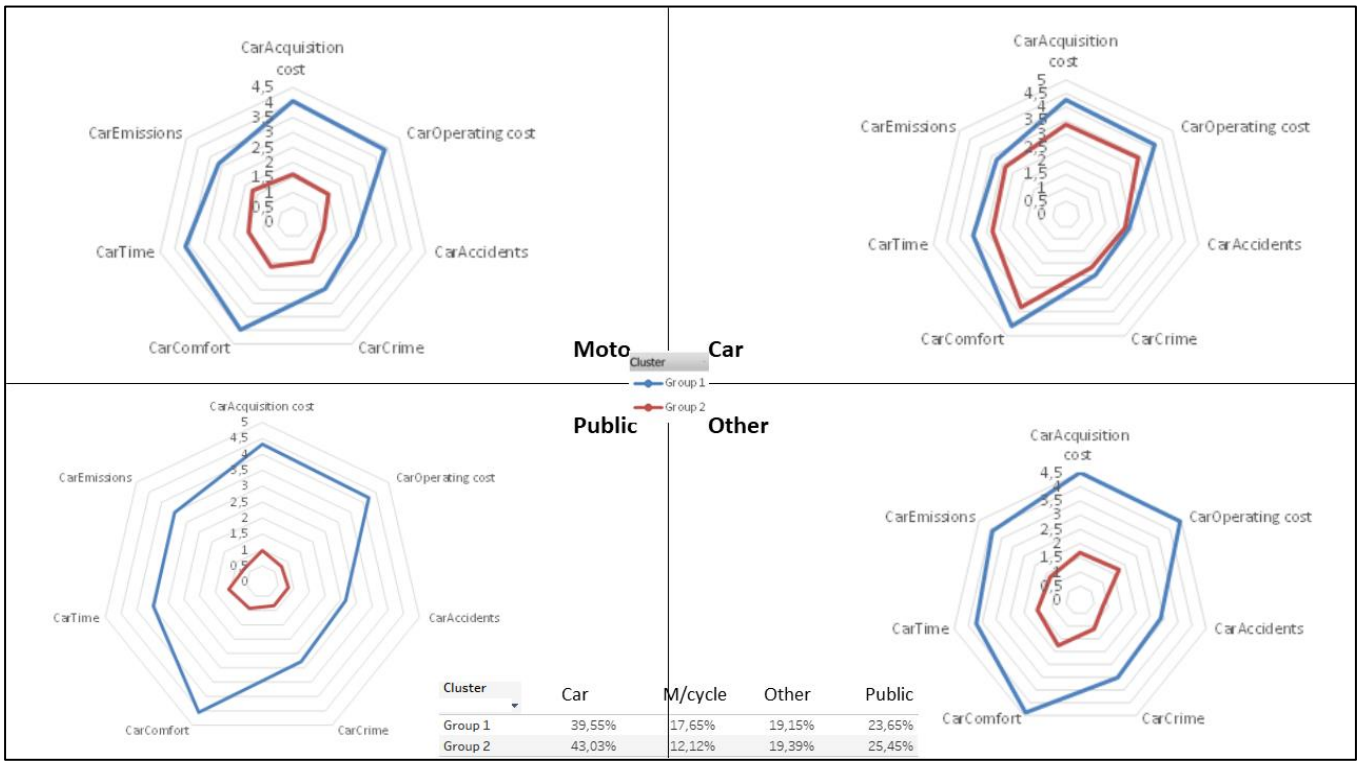

Figure 20. Average perception of car attributes differentiated by users' actual mode of transport.

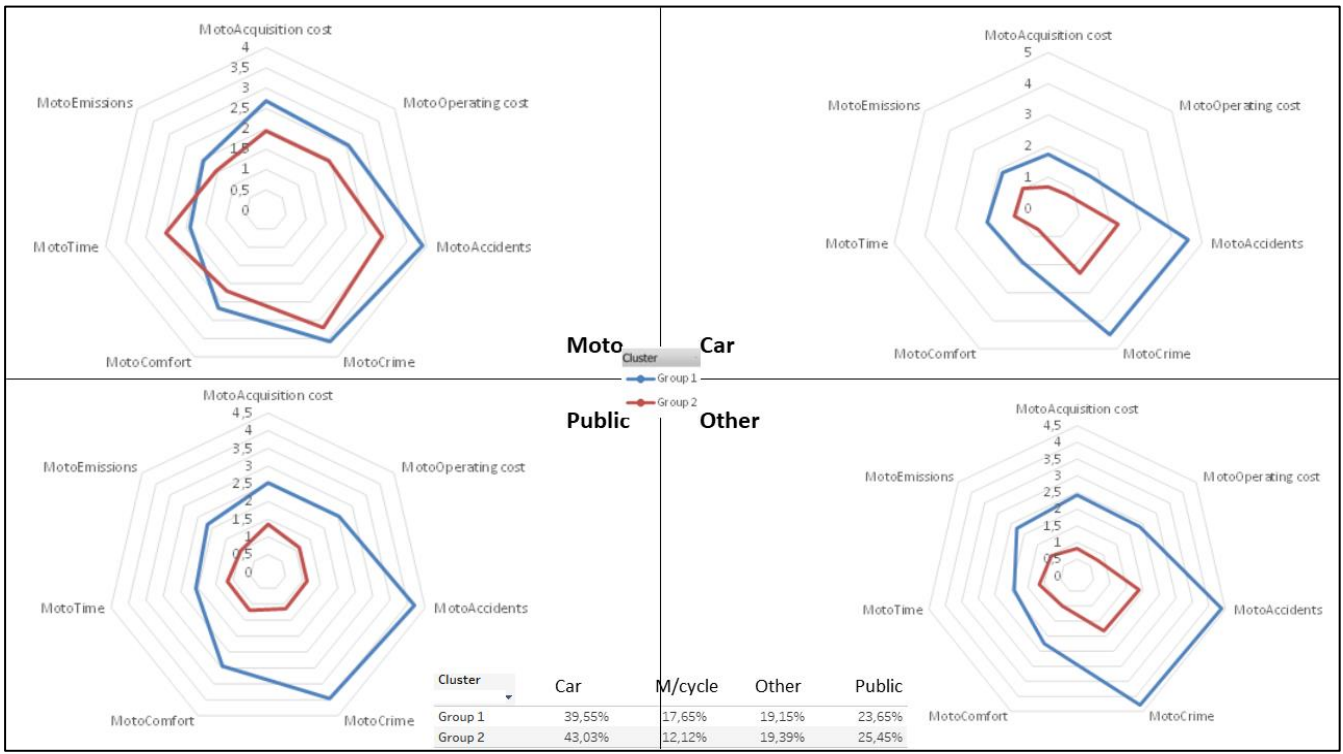

Figure 21. Average perception of motorcycle attributes differentiated by users' actual mode of transport.

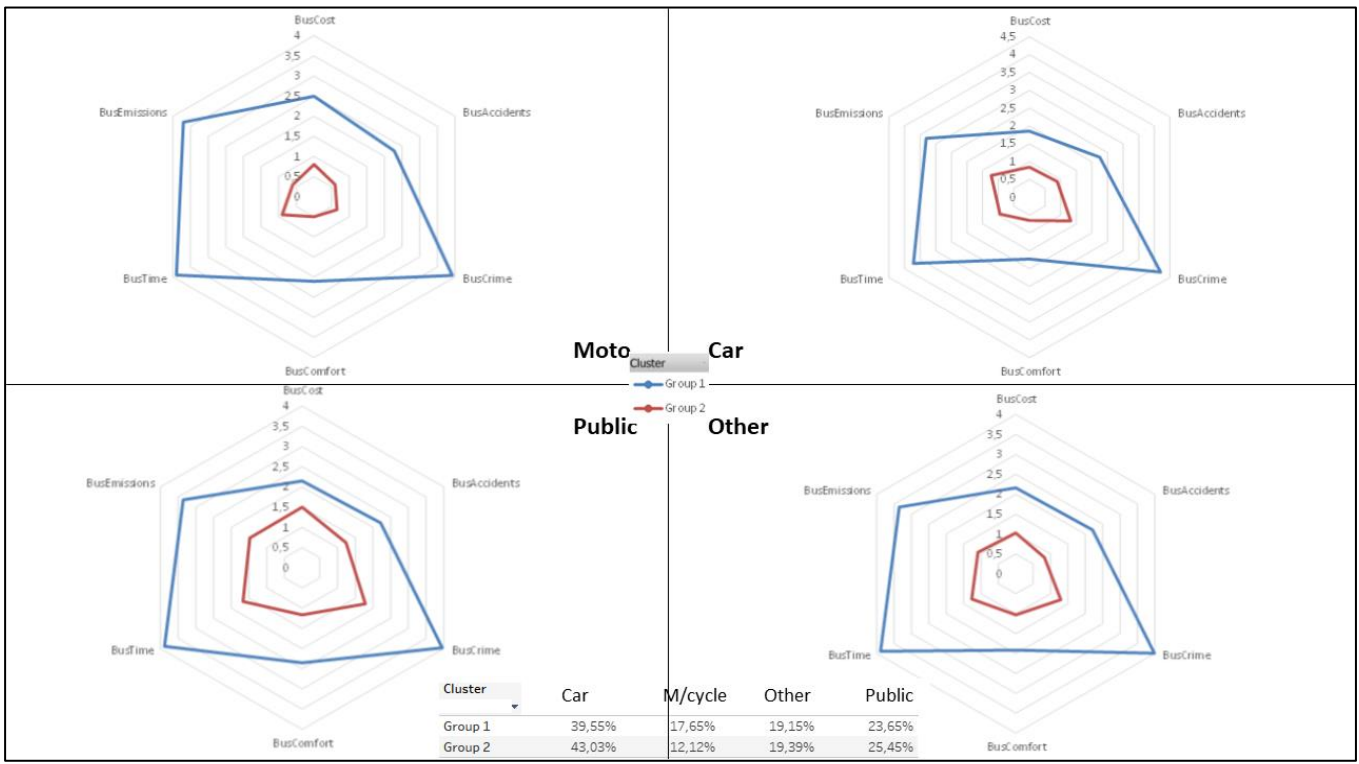

Figure 22. Average perception of public transport attributes differentiated by users' actual mode of transport.

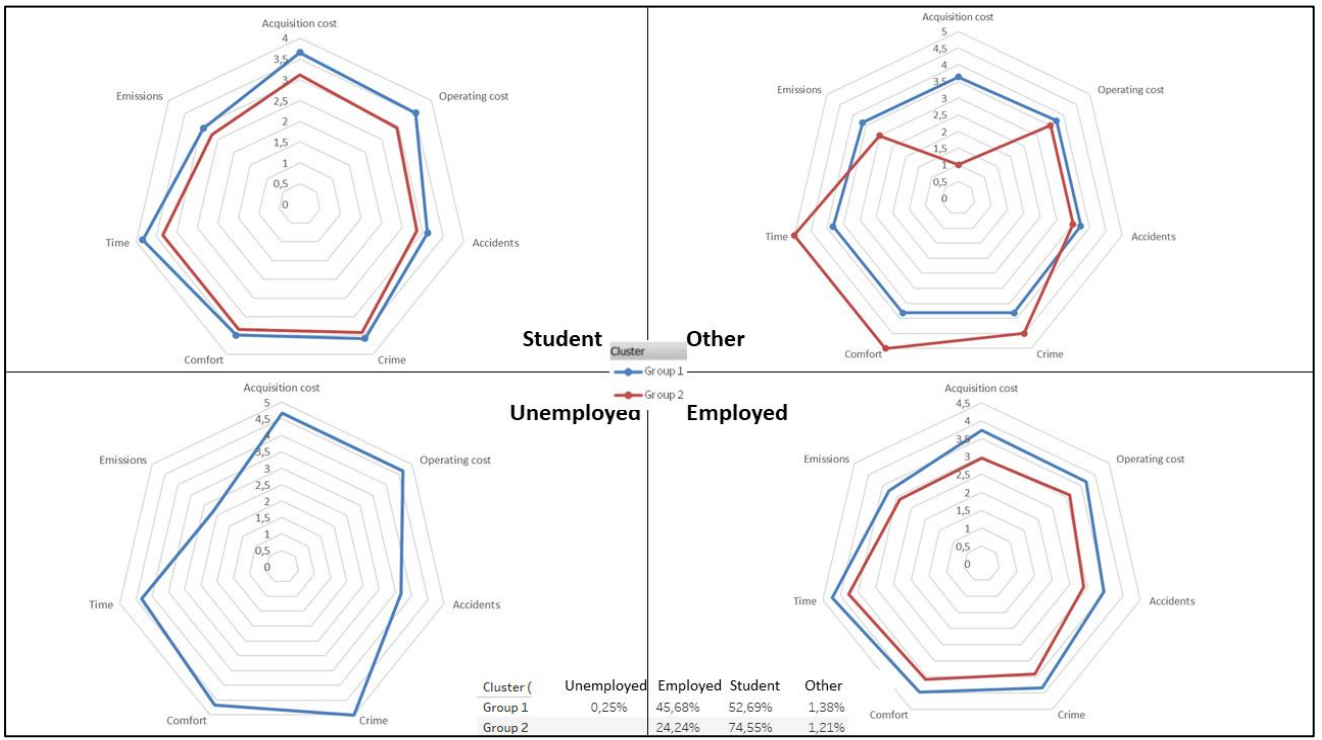

Figure 23. Factors influencing the choice of transportation mode differentiated by users' occupation.

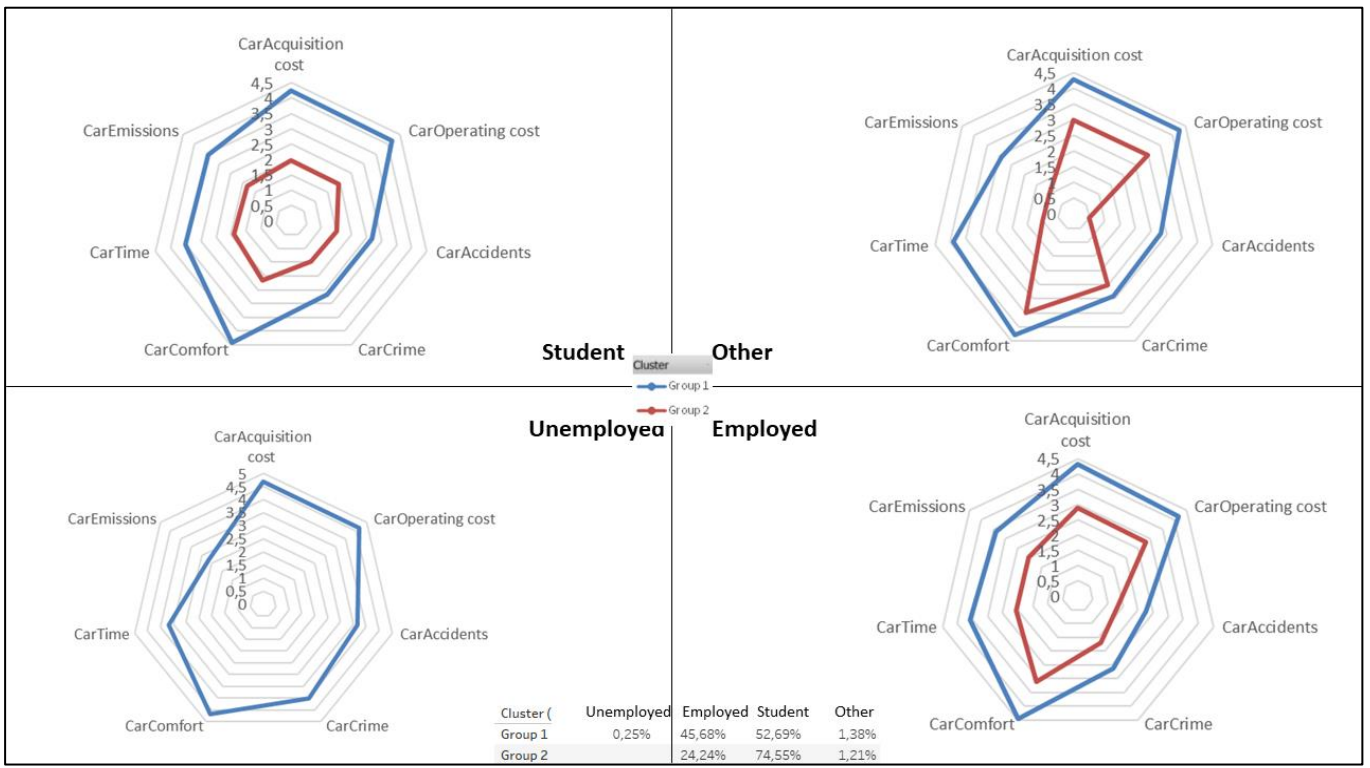

Figure 24. Average perception of car attributes differentiated by users' occupation.

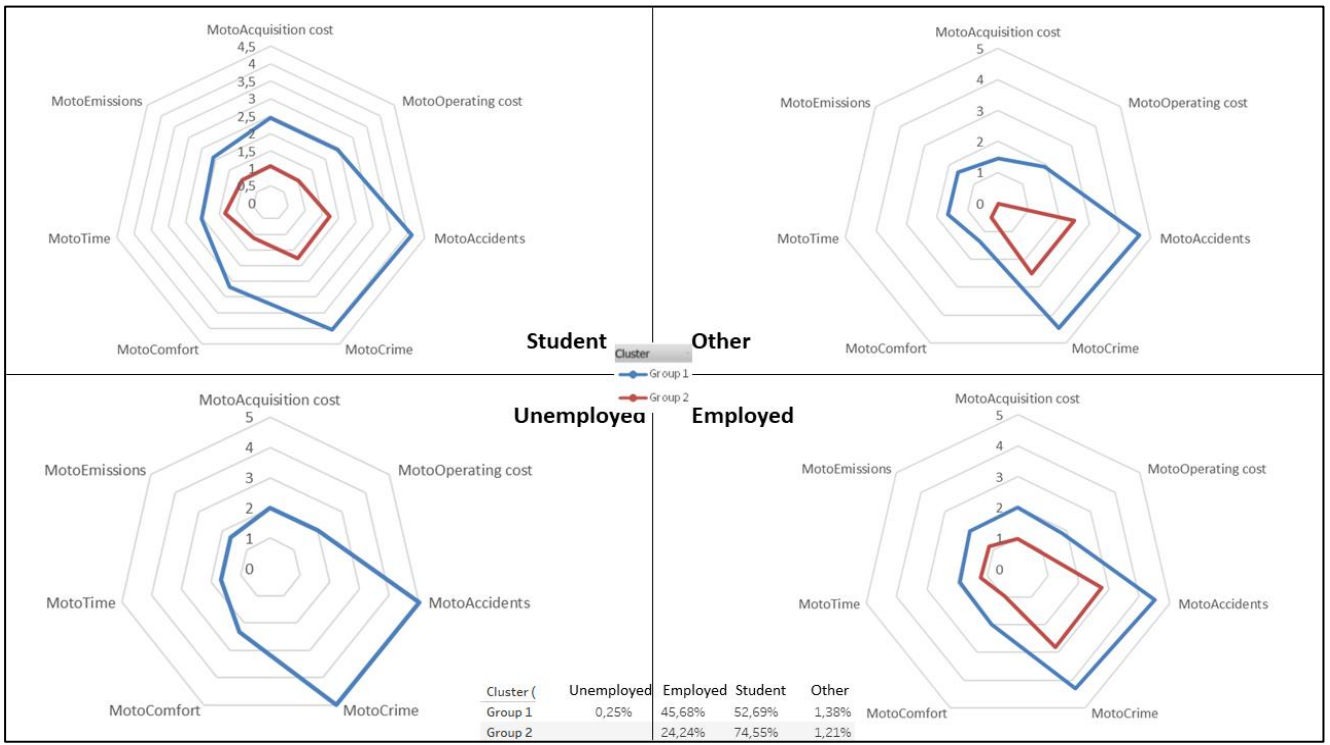

Figure 25. Average perception of motorcycle attributes differentiated by users' occupation.

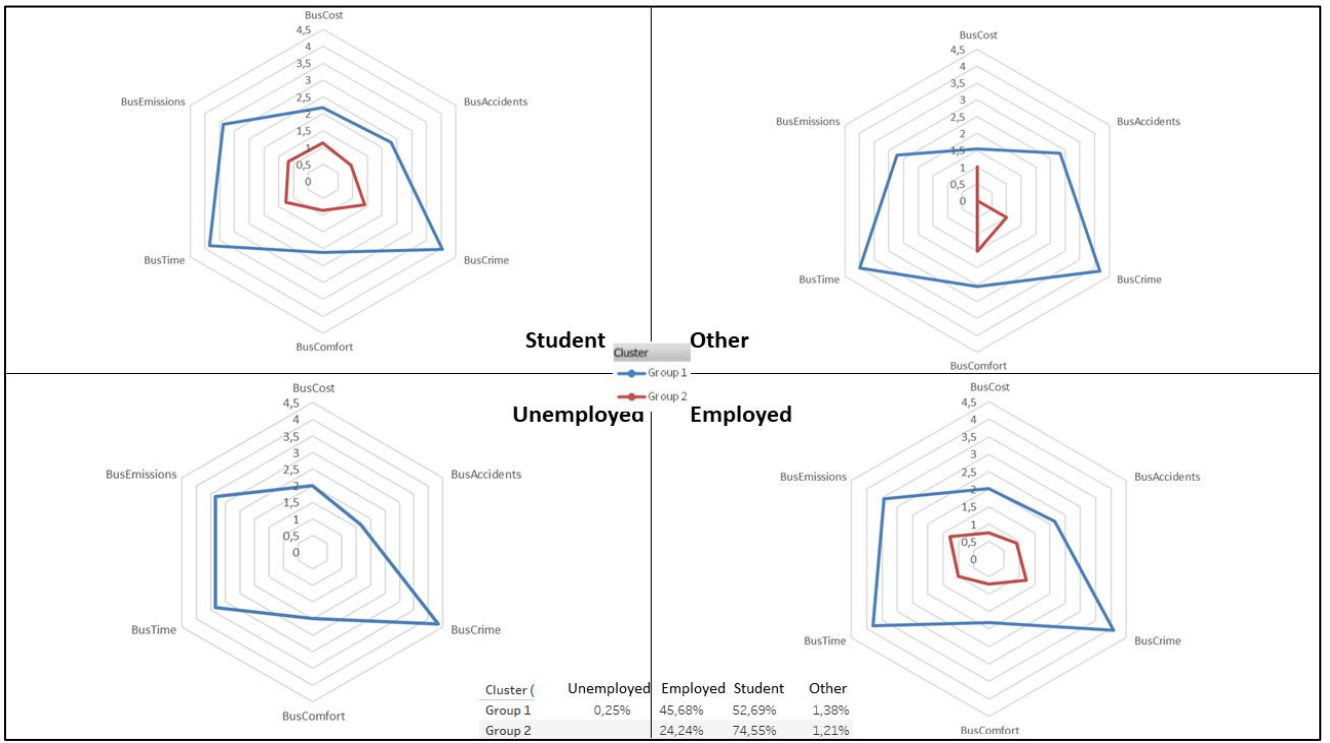

Figure 26. Average perception of public transport attributes differentiated by users' occupation.